\documentclass{aa}
\usepackage{natbib}
\usepackage{graphicx}
\usepackage{captcont}

\begin{document}
 
                          
\title{Kinematics of the Local Universe XIV.}
\subtitle{Measurements from the 21 cm line and the HI mass function from a homogeneous catalog gathered with the Nan\c cay radio telescope}

\author{G.\ Theureau\inst{1,2,3}  \and N.\ Coudreau\inst{3} \and N.\ Hallet\inst{3} 
\and M.\ 0.\ Hanski\inst{1,5} \and M.\ Poulain\inst{1}}
\institute{LPC2E, CNRS, Univ. Orl\'eans, F45071 Orleans Cedex 02, France
\and
Station de Radioastronomie de Nan\c cay, Observatoire de Paris, PSL Research University, CNRS, Univ. Orl\'eans, F18330 Nan\c cay, France
\and
LUTh, Observatoire de Paris, PSL Research University, CNRS, F92195 Meudon Principal Cedex, France
\and
GEPI, Observatoire de Paris, PSL Research University, CNRS, F92195 Meudon Principal Cedex, France
\and
Tuorla observatory, University of Turku, SF 21500 Piikki\"o, Finland
}
\date{Received : 2016 September 30 \hspace*{3cm} / accepted : 2016 November 4}

\abstract{}{This paper presents 828 new 21 cm neutral hydrogen line 
measurements carried out with the FORT receiver of the meridian 
transit Nan\c cay radio telescope (NRT) in the years 2000 -- 2007.}{ 
This observational program  was part of a larger project aimed at 
collecting an exhaustive and magnitude-complete HI extragalactic 
catalog for Tully-Fisher applications. Through five massive data releases, 
the KLUN series has collected a homogeneous sample of 4876 HI-spectra of spiral galaxies, 
complete down to a flux of 5 Jy.km.s$^{-1}$ and with declination $\delta > -40^{\circ}$.}{
We publish here the last release of the KLUN HI observational program, corresponding to the faint end of the survey,
with HI masses ranging from 5 10$^8$ to 5 10$^{10}$ solar masses. 
The size of this final sample is comparable to the catalogs  
based  on the Arecibo and Parkes radio telescope campaigns,  and it allows general HI mass distribution studies 
from a set of homogeneous radio measurements.}{}


\authorrunning{Theureau et al.}
\titlerunning{Kinematics of the Local Universe XIV}
 
\maketitle

\section{Introduction}

The present paper is the last in a long series of extragalactic HI spectra releases obtained from the 
long-term EDS \footnote{Extragalactic Distance Scale} and KLUN\footnote{for Kinematics in the Local UNiverse} surveys at the Nan\c cay radio telescope. 
We publish here a collection of new HI line measurements of spiral galaxies, complementing previous 
publications by   \citet{bot92},  \citet{bot93} (EDS), \citet{nel96},  \citet{the98a}, \citet{pat03b},
 \citet{the05},  and  \citet{the07} (KLUN). The KLUN program has received the label of key project 
 of the refurbished instrument (FORT receiver) and was allocated an average of 20 \% of the 
 observing time from mid 2000 to late 2006. 

This survey is complementary to other large HI projects led in the 2000s, such as the blind surveys
HIPASS\footnote{http://www.atnf.csiro.au/research/multibeam/release/} with the Parkes radio telescope
(\citealt{mey04}) and ALFALFA with the Arecibo radio telescope (\citealt{hay11}), 
or the Arecibo HI data compilation by \citet{spr05}.
In the last period of the survey corresponding to the present publication, 
the majority of the galaxies were observed from Nan\c cay in the range 
(-40$^{\circ}$,+0$^{\circ}$) in declination, favoring the declination range unreachable by Arecibo. 
Furthermore, our aim was to fill the gaps left in the last
Hyperleda HI compilation by \citet{pat03b} in order to reach
well-defined selection criteria in terms of redshift coverage and magnitude 
completeness for Tully-Fisher applications. The input catalog has been put together from a compilation of the Hyperleda extragalactic database
completed by the 2.7 million galaxy catalog extracted from the DSS (\citet{pat03a} ), and 
the releases of the DEep Near Infrared Survey  (DENIS; \citealt{pat05}) and the 2 
Micron All Sky Survey  (2MASS; \citealt{jar00}) near-infrared (NIR) CCD surveys. All targeted objects were inspected
by eye on optical and/or NIR images, to select only late-type galaxies, from Sa to Sm morphologies.

The total Nan\c cay HI catalog used for the KLUN analysis contains 4876 spiral galaxies, 
whose 21 cm line spectra were acquired across two receiver generations, and processed with a 
homogeneous reduction and calibration pipeline, all along the total observation campaign, 
from ~1992 to 2006 (see in particular \citet{the98a} and \citet{the05} for details).
This HI catalog is comparable in size with the Parkes and Arecibo catalogs, which were made public around 2005,
and allows  the same kind of general statistical HI mass studies as published respectively by  \citet{zwa05} 
and \citet{spr05}. We thus complement the data publication with a few results in terms of HI mass and its variation with galaxy morphology.

The present paper is structured in two main sections. The Nan\c cay radio telescope (NRT), the processing chain, the reduced HI 
data\footnote{Tables 2 and 3, together with HI profiles in ASCII format are available via anonymous ftp to cdsarc.u-strasbg.fr (130.79.128.5)
or via http://cdsweb.u-strasbg.fr/cgi-bin/qcat?J/A+A/},
 and the general properties of the catalog are presented in Section~\ref{SecObs}; 
we repeat there  the main characteristics of the instrument and of the data, which were described more extensively 
in  \citet{the05}. Section~\ref{SecHImass} shows  an example of HI mass function analysis, together with elements of comparison with previous extensive results 
by \citet{spr05}, \citet{zwa05}, and \citet{hay11}.

\begin{table}
\caption{Statistics of the detected galaxies vs. HI profile class.}
\label{TabQ}
\centering
\begin{tabular}{ccc}
\hline \hline
 profile class & no. of gal. & no. of HI confusions \\
\hline
A &  160  & 11\\
B & 303 & 34 \\
C & 174  & 46 \\
D &  86  & 5 \\
E & 104  & 3 \\
\hline
\end{tabular}
\end{table}

\begin{figure}
\resizebox{\hsize}{!}{\includegraphics{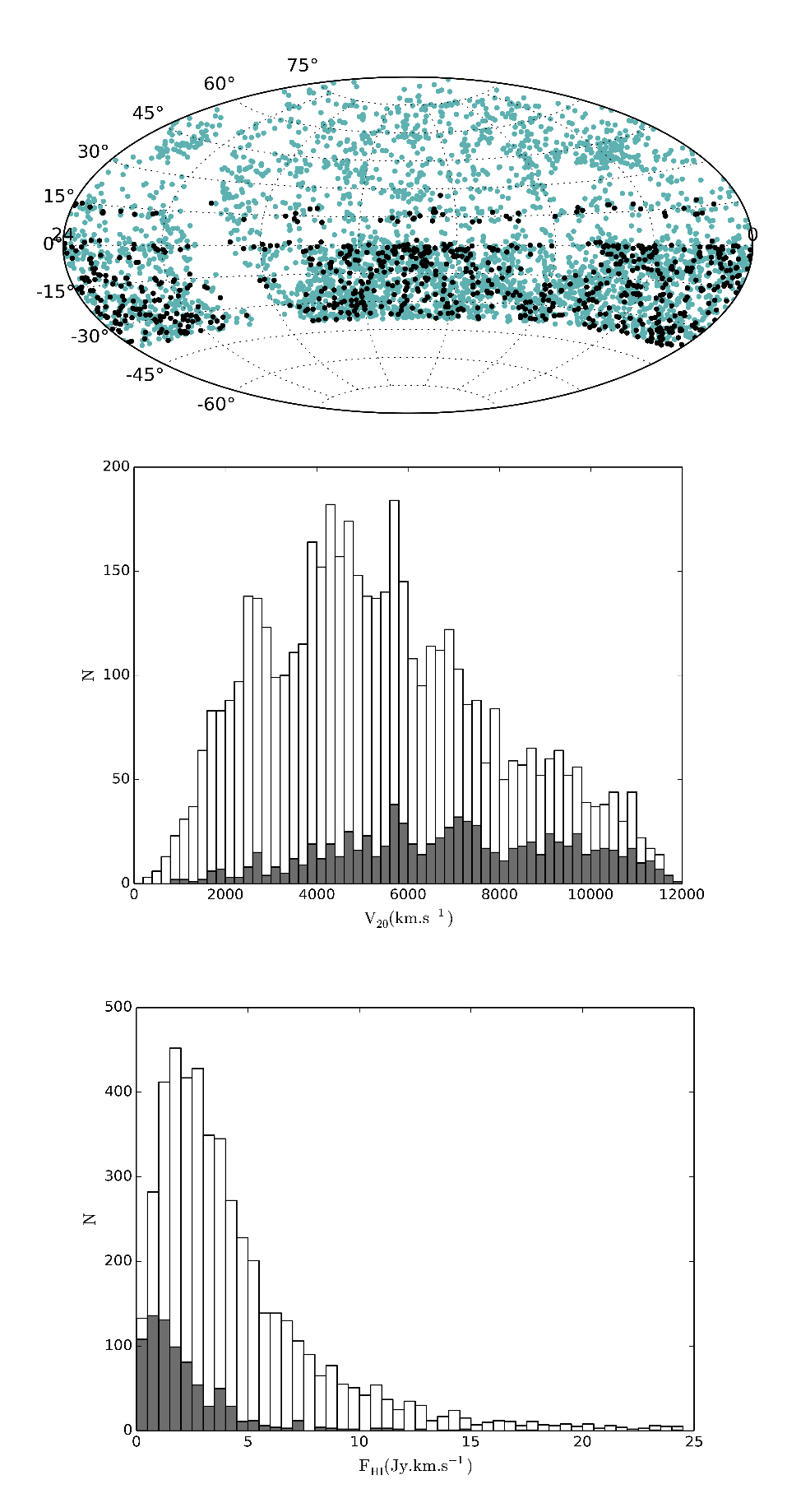}}
\caption{General statistics of the KLUN HI catalog. Top : Aitoff projection of the KLUN composite Nan\c cay HI-data catalog (cyan dots: \citet{the98a}, 
\citet{pat03b}, \citet{the05}, \citet{the07}; black dots: the galaxies added in this publication).
Middle: distribution of radial velocities (optical convention) for the KLUN composite catalog (white columns) and this paper (gray columns).
Bottom:  distribution of beam-corrected HI-fluxes (same samples).}
\label{FigStat}
\end{figure}

\begin{figure*}
\centering
\includegraphics[width=16.5cm]{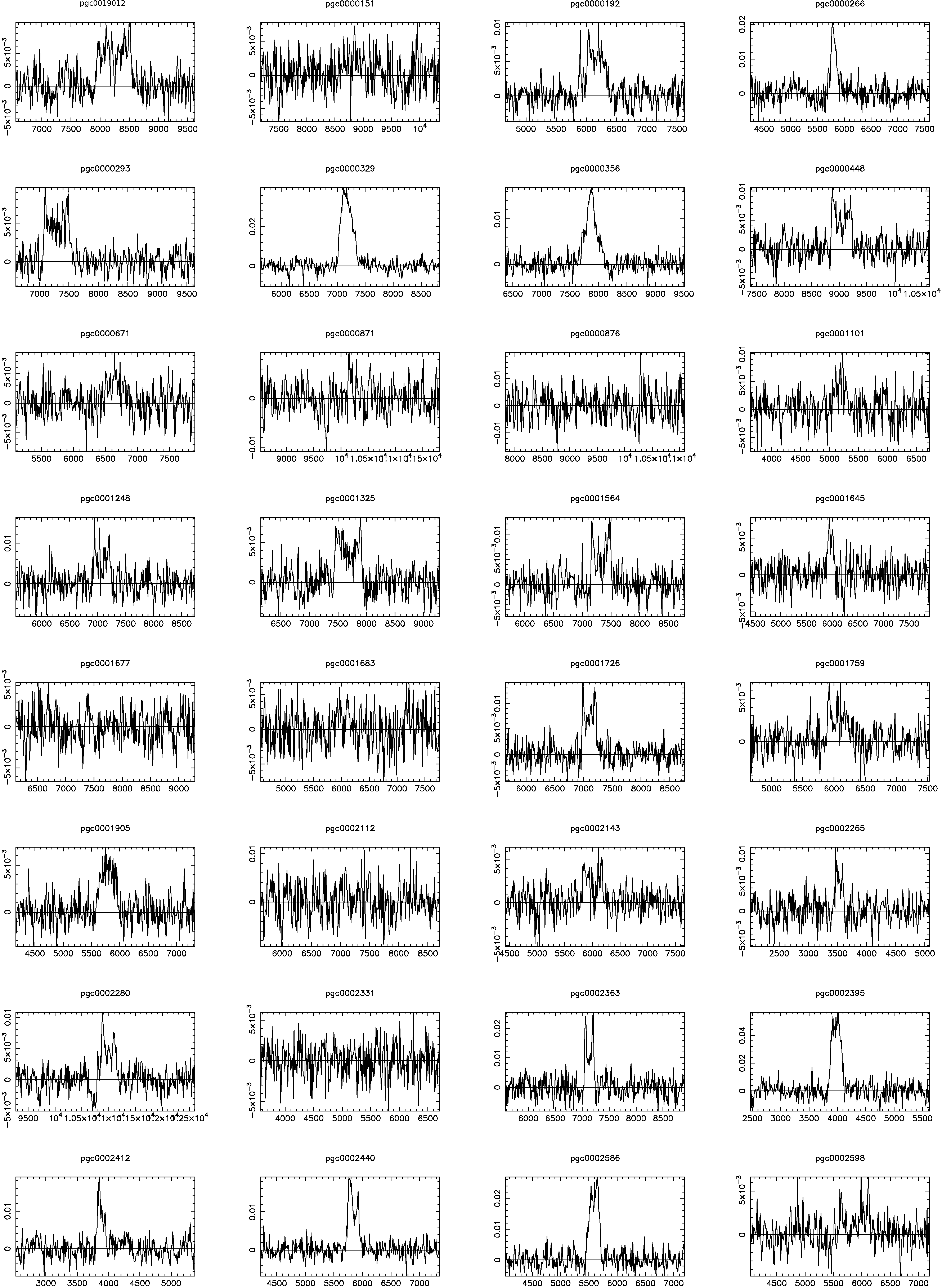}
\caption{{\bf -excerpt. }21 cm line profiles of galaxies listed in Table \ref{TabPar}. The  profiles 
are classified according to their PGC name which 
is written above each panel. Ordinate and abscissae axes are 
graduated respectively in km s$^{-1}$ and Jy. The heliocentric
radial velocities are expressed in terms of optical redshift 
$c \frac{\Delta \lambda}{\lambda}$. The horizontal line
represents the baseline of the profile, i.e. the zero flux level, from which the maximum
is estimated. Figure 2 continues in appendix A1.} 
\label{FigHI}
\end{figure*}

\begin{table*}
\small{
\setlength{\tabcolsep}{0.03in}
\begin{tabular}{lllrrrrrrrrrrrrrc}
pgc/leda   &    NAME      & RA (2000) DEC &  $V_{20}$  &  $\sigma_V$  &  $W_{20}$ &  
$W_{20c}$ & $\sigma_{W20}$ &  $W_{50}$ & $W_{50c}$ & $\sigma_{W50}$ &  F(HI) & F(HI)$_{c}$ & $\sigma_F$ &  $S/N$ & rms & Q \\
\hline
pgc0019012 & 2MASJ06252 & J062529.4-321701 &  8076.0 & 24.0 & 314.0 & 260.0 & 73.0 & 289.0 & 261.0 & 49.0 & 1.31 &  &  0.74 & 2.0 & 2.7 & B? \\
pgc0000151 & ESO349-020 & J000203.9-332802 &  & & & & & & & & & & & & 2.8 & E  \\
pgc0000192 &  PGC000192 & J000248.6-033622 &  6176.0  & 22.0 & 381.0 & 327.0 &  65.0 & 318.0 & 290.0  & 43.0  & 1.98 & 1.98 & 0.54 & 3.6 & 1.7 & Cc \\    
pgc0000266 &  PGC000266 & J000345.1-144140 &  5815.0  & 12.0 & 175.0 & 128.0 &  37.0 & 109.0 &   88.0  & 25.0  &  2.14 &  & 0.52 & 6.5 & 2.7 & B \\
pgc0000293 &  PGC000293 & J000412.1-143125 &  7307.0  & 17.0 & 478.0 & 424.0 &  51.0 & 430.0 & 401.0  & 34.0  & 2.22 & 2.28 & 0.49 & 4.0 & 1.4 & B  \\
pgc0000329 &  PGC000329 & J000447.2-013413 &  7182.0  &  5.0 & 323.0 & 269.0 &  15.0 & 280.0 & 252.0  & 10.0  & 8.51 & 8.56 & 0.72 & 13.3 & 2.6 & C? \\
pgc0000356 &  PGC000356 & J000507.6-034822 &  7884.0  & 15.0 & 357.0 & 303.0 &  44.0 & 169.0 & 142.0  & 29.0  & 3.28 &  3.3 & 0.42 & 9.2 & 1.6 & Cc \\
pgc0000448 & ESO409-018 & J000604.1-303743 &  9047.0  &  9.0 & 382.0 & 328.0 &  26.0 & 373.0 & 344.0  & 17.0  & 2.06 & 2.12 & 0.6 & 3.5 & 2.0 & A \\
pgc0000671 & ESO349-034 & J000924.4-364258 &  6613.0  & 14.0 & 243.0 & 190.0 &  41.0 & 237.0 & 208.0  & 27.0  &  0.8 &  0.8 & 0.6 & 1.8 & 2.6 & D \\
pgc0000871 &  PGC000871 & J001301.9-131526 & 10261.0  & 33.0 & 275.0 & 221.0 &  98.0 & 253.0 & 224.0  & 65.0  &  0.6 &  0.6 & 0.65 & 1.4 & 3.2 & D \\
pgc0000876 &  PGC000876 & J001303.4-304314 &  & & & & & & & & & & & & 5.8 & E \\
pgc0001101 &  PGC001101 & J001641.5-103311 &  5186.0  & 21.0 & 274.0 & 220.0 &  62.0 & 250.0 & 221.0  & 41.0  & 1.06 & 1.07 & 0.62 & 2.3 & 2.7 & D \\
pgc0001248 &  PGC001248 & J001920.8-105639 &  7073.0  & 21.0 & 345.0 & 291.0 &  62.0 & 297.0 & 268.0  & 41.0  & 1.94 & 1.97 & 0.73 & 3.3 & 2.9 & B \\
pgc0001325 & ESO473-016 & J002042.1-231658 &  7681.0  & 15.0 & 489.0 & 435.0 &  45.0 & 463.0 & 435.0  & 30.0  & 1.98 & 2.04 & 0.56 & 3.3 & 1.7 & B \\
pgc0001564 &  PGC001564 & J002518.3-143327 &  7327.0  &  7.0 & 349.0 & 295.0 &  22.0 & 339.0 & 310.0  & 15.0  & 2.11 & 2.14 & 0.61 & 4.2 & 2.4 & A \\
pgc0001645 &  PGC001645 & J002638.3-303301 &  5971.0  & 13.0 & 164.0 & 118.0 &  40.0 & 148.0 & 122.0  & 27.0 & 0.67 & 0.68 & 0.35 & 3.0 & 1.8 & A \\
pgc0001677 &  PGC001677 & J002714.6-074713 &  & & & & & & & & & & & & 2.2 & E \\
pgc0001683 &  PGC001683 & J002720.0-045329 &  & & & & & & & & & & & & 2.8 & E \\
pgc0001726 &  PGC001726 & J002802.4-080716 &  7101.0  &  8.0 & 263.0 & 209.0 &  25.0 & 242.0 & 214.0  & 16.0  & 2.05 & 2.05 & 0.48 & 5.4 & 2.0 & A? \\
pgc0001759 &  IC0018 & J002835.0-113512 &  5986.0  & 12.0 & 195.0 & 145.0 &  35.0 & 184.0 & 156.0  & 24.0 &  0.58 & 0.59 & 0.33 & 2.8 & 1.7 & C? \\
pgc0001905 &  IC0025 & J003112.1-002426 &  5773.0  & 18.0 & 375.0 & 321.0 &  55.0 & 345.0 & 317.0  & 37.0 &  1.42 & 1.42 & 0.46 & 2.9 & 1.4 & B \\
pgc0002112 & ESO294-023 & J003513.1-373329 &  & & & & & & & & & & & & 3.7 & E  \\
pgc0002143 & NGC0166 & J003548.8-133638 &  6007.0 &  14.0 & 371.0 & 317.0 &  41.0 & 362.0 & 333.0  & 27.0 &  0.93 & 0.95 & 0.44 & 2.2 & 1.5 & C \\
pgc0002265 & ESO411-002 & J003755.2-285523 &  3530.0  & 14.0 & 170.0 & 123.0  & 41.0 & 147.0 & 121.0  & 27.0  & 0.85 & 0.87 & 0.4 & 3.5 & 2.1 & B \\
pgc0002280 &  PGC002280 & J003818.4-145056 & 10983.0  & 19.0 & 383.0 & 329.0  & 57.0 & 349.0 & 320.0  & 38.0  & 1.68 & 1.68 & 0.54 & 3.0 & 1.7 & C? \\
pgc0002331 &  NGC0191 & J003859.4-090010 &  & & & & & & & & & & & & 2.2 & E  \\
pgc0002363 & ESO411-004 & J003918.8-295644 &  7130.0  &  5.0 & 182.0 & 134.0  & 15.0 & 170.0 & 143.0  & 10.0  & 2.42 & 2.45 & 0.59 & 6.9 & 3.2 & A \\
pgc0002395 & NGC0207 & J003940.7-141414 &  3982.0 &   8.0 & 253.0 & 200.0 &  24.0 & 208.0 & 179.0  & 16.0 &  9.69 & 9.75 & 1.39 & 8.2 & 5.3 & A? \\
pgc0002412 &  PGC002412 & J004005.6-200349 &  3876.0  & 10.0 & 165.0 & 119.0  & 30.0 & 136.0 & 111.0  & 20.0  & 1.52 & 1.54 & 0.47 & 5.4 & 2.5 & B \\
pgc0002440 & IC1571 & J004037.9-001950 &  5836.0 &   7.0 & 241.0 & 188.0 &  21.0 & 200.0 & 172.0  & 14.0 & 2.81 & 2.87 & 0.43 & 9.0 & 1.9 & A? \\
pgc0002586 &  PGC002586 & J004317.7-063834 &  5583.0  &  7.0 & 250.0 & 197.0  & 21.0 & 212.0 & 183.0  & 14.0  &  4.3 & 4.32 & 0.62 & 8.7 & 2.5 & A? \\
pgc0002598 &  PGC002598 & J004328.5-062055 &  5860.0  & 36.0 & 574.0 & 520.0 & 107.0 & 536.0 & 507.0  & 71.0  & 0.94 & 0.94 & 0.52 & 1.7 & 1.6 & D  \\
  ...    &    ...       &    ...             &   ...  &  ... &   ... &   ... &   ... & ...   & ...   & ...   & ...   &  ...   & ...  & ...  & ... & ...  \\\hline
\end{tabular}
}
\caption[]
{{\bf -excerpt.} Astrophysical HI-parameters (first 32 galaxies corresponding to Fig. \ref{FigHI}).  \\
Column 1: PGC or LEDA galaxy name; \\
Column 2: most usual galaxy name; \\
Column 3:  J2000 equatorial coordinates; \\
Column 4: systemic heliocentric radial velocity (km s$^{-1}$); \\
Column 5: rms error (km s$^{-1}$); \\
Column 6: total line width at 20\% of the maximum intensity (km s$^{-1}$); \\
Column 7: total corrected line width at 20\% (km s$^{-1}$); \\
Column 8: rms error (km s$^{-1}$); \\
Column 9: total line width at 50\% of the maximum intensity (km s$^{-1}$); \\
Column 10: total corrected line width at 50\% (km s$^{-1}$); \\
Column 11: rms error (km s$^{-1}$); \\
Column 12: observed HI-flux (Jy km s$^{-1}$); \\
Column 13: beam-filling corrected HI-flux (Jy km s$^{-1}$); \\
Column 14: rms error (Jy km s$^{-1}$); \\
Column 15: signal-to-noise ratio; \\
Column 16: rms noise; \\ 
Column 17: quality code (\ref{SecQ}) ; flag (``c'' indicates confirmed HI confusion with the emission of another galaxy; ``?'' means that confusion is suspected but not certain).}
\label{TabPar}
\end{table*}

\begin{table*}
\small{
\begin{tabular}{llll}
pgc/leda  &  Q & Type & comments \\
\hline
pgc0019012 & B? & Sa &=ESO426-007=2MASXJ06252935-3217012 \\ 
&&& poss HI confusion w pgc3081242 in off beam V=8284 \\
pgc0000151 & E  & Sa  &  =ESO349-020 V=8760 \\
pgc0000192 & Cc & Sc  & HI confusion w pgc0000176 3 arcmin N-NW V=6465 Sbc \\
&&& and pgc0000183 3 arcmin W, V=6290 prob late-type \\
pgc0000266 & B  & Sc &  \\
pgc0000293 & B  & Sc & \\
pgc0000329 & C? & Sab & poss HI confusion w pgc0000330 4 arcmin N V=7110 \\
&&& and pgc0000352 1 arcmin S-SE, V=7187 both prob late-type \\
pgc0000356 & Cc & Sc  & HI confusion w pgc169982 3 arcmin SW V=7851 face-on \\
pgc0000448 & A  & Sab & \\
pgc0000671 & D & Sc  & =ESO349-034 \\
pgc0000871 & D  & S0-a & V=10330 \\
pgc0000876 & E  & S0-a & V=9320, pgc0713877 in beam w V=9500 \\
pgc0001101 & D  & S0-a & V=5196 \\
pgc0001248 & B  & Sab &  \\
pgc0001325 & B & Sa  & =ESO473-016 \\
pgc0001564 & A  & Sc & \\
pgc0001645 & A  & Sa & \\
pgc0001677 & E  & S0-a & V=6728 cataloged as late-type in Huchra et al. 2012 ApJS 199, 26 \\
pgc0001683 & E  & Sc  & V=6123 \\
pgc0001726 & A? & Irr? & V=116 in Lavaux \& Hudson 2011 MNRAS 416, 2840; \\
&&& poss HI confusion w pgc172103, 6 arcmin SW V=7217 \\
pgc0001759 & C? & Sb  & =IC0018, merger ? poss HI confusion w pgc138190 in off beam V=6075 \\
&&& and w IC0019, 3.5 arcmin S V=6177 \\
pgc0001905 & B  & Sab  & =IC0025 \\
pgc0002112 & E  & Sa  & =ESO294-023 V=7242 \\
pgc0002143 & C  & Sa  & =NGC166 \\
pgc0002265 &B  & Sc  & =ESO411-002 \\
pgc0002280 & C? & Sab & poss HI confusion w group of 2MASS gal in off beam at J003645.8-145953 no V \\
pgc0002331 & E  & Sc  & =NGC0191 V=6076 \\
pgc0002363 & A  & Sc  & =ESO411-004 \\
pgc0002395 & A? & Sa &  =NGC0207, pgc138206 3.5 arcmin N V=3959, prob early-type, confusion ? \\
pgc0002412 & B  & Sc & \\
pgc0002440 & A? & Sb &  =IC1571, poss HI confusion w nearby group within 4 arcmin and <V>=5800 \\
pgc0002586 & A? & Sc  & poss HI confusion w pgc2583 1.5 arcmin S no V Sm \\
pgc0002598 & D  & S0  & low S/N \\
 ...  & ...  & ... & ... \\
\hline 
\end{tabular}
}
\caption[]
{{\bf -excerpt.} Notes on HI-observations (first 32 galaxies corresponding to Fig. \ref{FigHI}).  \\
Column 1: PGC or LEDA galaxy name; \\
Column 2: quality code and HI-confusion flag ``c'' (confirmed) or ``?'' (possible); \\
Column 3: comments;  comp=companion, cf=comparison field, poss=possible, w=with,  prob=probable. }
\label{TabCom}
\end{table*}

\section{ HI campaign} 
\label{SecObs}

\subsection{Sample characteristics}

As a last KLUN HI data release, we present here 21 cm line measurements obtained for 828 targeted galaxies, 
observed between mid 2000 and late 2006, the majority of which were close to the limit of detection. 
With this publication, the total contribution of the so-called KLUN program to the HI spectra budget
amounts to 4876 galaxies, collected over a period of $\sim$ 15 years, with a homogeneous data reduction pipeline.
This composite HI sample was built by gathering the data from the five publications of the KLUN series: 
 \citet{the98a}, \citet{pat03b}, \citet{the05}, \citet{the07}, and this paper.

Figure\ \ref{FigStat} shows the general statistics of the Nan\c cay HI catalog for both the total KLUN composite sample and 
the subpart of 828 new measurements presented in this paper. As mentioned above, the vast majority of the targets were chosen 
outside the Arecibo field, between the Nan\c cay declination lower limit -40$^{\circ}$ and the Arecibo threshold at -1$^{\circ}$. 
Most of the observed galaxies are in the range 4\,000-10\,000 km.s$^{-1}$, where the lack of 
Tully-Fisher measurements in the literature was the most critical. 
Since this publication presents the epilog of the KLUN observational program, a large number of the HI spectra released here
are from faint HI galaxies, close to the limit of detection (with HI-line S/N often below 5) and which often required several hours 
of exposure time to get measurable line parameters. 
The NRT is a quasi-meridian instrument, with a maximum one-hour daily window at a given RA, which means that the  observations 
of a given galaxy were systematically scheduled to spread over several transits, sometimes separated by months. Actually, the integration time per 
galaxy and per polarization spans from 15 minutes  (e.g.,  pgc36297) to 6 hours (e.g., pgc4566), with an average value around 2 hours. 
The final spectrum is obtained by summing the two banks in vertical and horizontal  polarizations and applying a boxcar smoothing to get the
final resolution of typically $\sim$10 km.s$^{-1}$ at 21 cm. During the period covered by the survey,
the evolving S/N was checked   after each transit and a given target was classified as ``completed''  only once it reached S/N~$\geq 5$, 
or if we reached a rms noise smaller than 2 mJy, which is a reasonable detection threshold for the instrument.  
For these faint objects it was generally the latter case, since most of the high S/N HI galaxy spectra were published
in the previous KLUN releases. 

\subsection{Measured HI parameters}

The NRT design, characteristics of the FORT receiver, spectrograph, and data pipeline were exhaustively 
described in  \citet{the05}.

Here we only describe the corrections applied to the raw measurements of the HI-line to get the 
parameters of astrophysical interest. 
An excerpt of the catalog of HI-astrophysical parameters, of observers' 
comments, and plots of the 21 cm line spectra is given respectively in Table \ref{TabPar}, Table \ref{TabCom}, and Fig \ref{FigHI}.
The full sample is available in electronic format at the CDS$^4$.

\paragraph{Spectrum quality code and HI confusion check.}
\label{SecQ}

Table \ref{TabPar} contains all the reduced HI parameters. The local environment and possible HI confusion 
within the source beam or off beam has been checked by eye on the basis of online DSS and SDSS images and 
available astrophysical data from the Hyperleda and NED extragalactic data bases. Table~\ref{TabCom} provides corresponding
comments, when necessary, for each galaxy. Comments mainly concern  object designation, peculiar morphology,
peculiar HI line shape, spectrum quality, or HI confusion. The spectra and extracted data are assigned 
a quality code from A to E, with decreasing spectrum quality (see \citet{the05} for details). 
In short, A and B classes correspond to well-defined HI profiles and subsequent measured HI parameters, while in classes C and D
only the radial velocity is reliable owing to disrupted baseline and/or a signal-to-noise ratio (S/N) that is too low, and E stands for non-detection. 
In addition, a flag of ``?'' or ``c'' indicates suspected or confirmed HI line confusion. 
The distribution of the targets among the different classes is summarized in Table \ref{TabQ}, together with the respective number of HI confusion.

\paragraph{Radial velocities.} 
\label{RadVel}

Our observed radial velocities are listed in Table \ref{TabPar} (column~4)
and correspond to the median point of the 21 cm line profile measured
at 20\% of maximum intensity. The internal mean error on $V_{20}$ is calculated according to
\citet{fou90} as 
\[ \sigma(V_{20}) = \frac{4 \cdot (R \cdot \alpha)^{1/2}}{S/N}, \]
where $R$ is the actual spectral resolution, $\alpha = 
(W_{20}-W_{50})/2$ is the slope of
the line profile, and $S/N$ is the signal-to-noise ratio.
The average of $\sigma (V_{20})$ is about 8 km.s$^{-1}$.
The radial velocity distribution of the sample is shown in Fig. \ref{FigStat} and is compared with the whole Nan\c cay KLUN catalog.

\paragraph{Line widths} 
\label{LinWid}

are measured on the observed profile at two standard
levels corresponding to 20\% and 50\% of the maximum intensity 
of the line. The results listed in Table \ref{TabPar}, columns 6 and 9, have been corrected to 
the optical velocity scale. We also provide line 
widths corrected for resolution effect (\citet{fou90}) in columns 7 and 10. 
The mean measurement error is taken equal to $3 \cdot \sigma(V_{20})$ and 
$2 \cdot \sigma(V_{20})$ for the 20\% and 50\% widths, respectively. 
The data presented here are not corrected for internal 
velocity dispersion. Details about these corrections can be found in 
\citet{bot90}, \citet{fou90}, or  \citet{pat03b}.

\paragraph{HI-fluxes} 
\label{HIflu}

are calibrated using as templates a set of well-defined radio continuum sources observed each month 
and further checked with a sample of ``typical'' galaxies observed regularly   during the survey period and reduced through the same pipeline.
Again, the detailed description of the flux calibration is given in \citet{the05}.
HI-fluxes $F_{HI}$ (Table \ref{TabPar}, column 12) are expressed in Jy km.s$^{-1}$.
The values given in column 13 are corrected for beam-filling according to \citet{pat03b}
\[ F_{HIc} = B_f \cdot F_{HI}, \]
where $F_{HI}$ is the observed raw HI-flux, 
\[ B_f = \sqrt{(1+xT)(1+xt)}, \]
\[ T= (a_{25}^2 \sin^2\beta + b_{25}^2 \cos^2\beta)/\theta_{EW}^2, \]
\[ t= (a_{25}^2 \cos^2\beta + b_{25}^2 \sin^2\beta)/\theta_{NS}^2, \]
$\theta_{EW}$ and $\theta_{NS}$ are the half-power beam dimensions
of the Nan\c cay antenna, $\beta$ is the position angle of the galaxy defined  northeastwards, and
$a_{25}$ and $b_{25}$ are respectively the photometric major and minor axis. The parameter x 
is $x=0.72 \pm 0.06$ \citep{bot90}. 
Finally, owing to the Kraus design of the telescope, the position angle of the beam is tilted when the focal 
chariot is moving away from the meridian. For the NRT, the correct expression for the PA is  
$ PA  = \arcsin(\sin(AH) \times \sin\delta)$, where AH is the hour angle. 
The AH from the meridian transit being less than 30 min, the tilt never 
exceeds 7.5$^{\circ}$ (or even 4.8$^{\circ}$ in our case given 
the actual distribution of our sources in declination). This results in a negligible effect on the average beam shape
during a single meridian transit.

\begin{figure*}
\centering
\includegraphics[width=16.5cm]{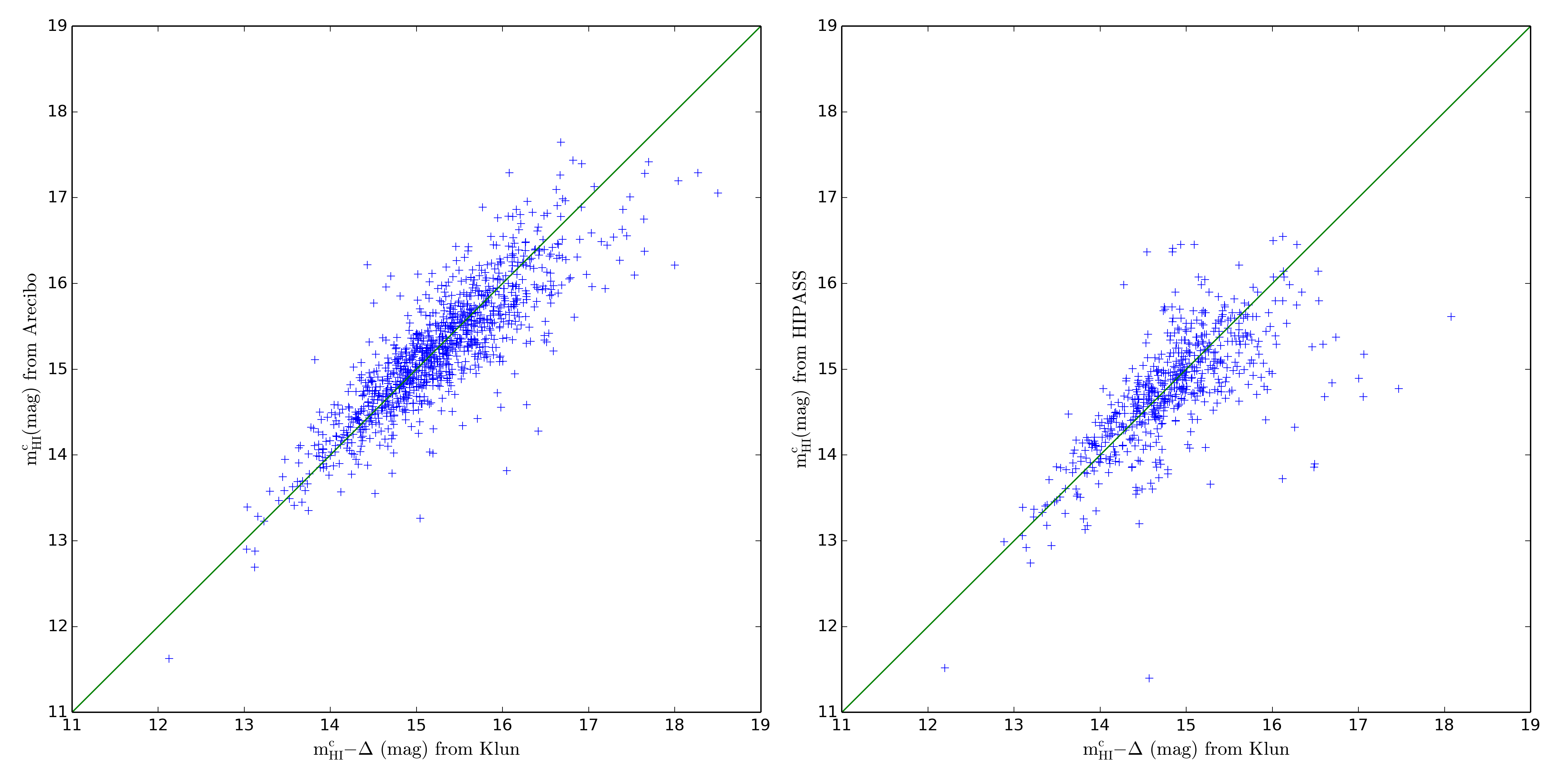}
\caption{Comparison of some of our HI-flux values with some independent measurements from  \citet{spr05} and HIPASS (\citealt{mey04}).
Fluxes are expressed in magnitude $m_{\rm{HI}}$ scale, with $m_{\rm{HI}} = -2.5 \rm{log}(0.2366F_{\rm{HI}}) + 15.84$,
as in \citet{pat03b}. The green line shows the first diagonal.}
\label{FigComp}
\end{figure*}

The HI flux calibration is homogeneous among the KLUN series. 
These measurements, however, need to be homogenized with the data from other telescope catalogs before a proper comparison can be carried out.  
For the HI mass function calculations, we used a shift of respectively $\Delta m_{\rm{HI}}$ = 0.27 and $\Delta m_{\rm{HI}}$ = 0.20 
in magnitude scale between the KLUN sample $F_{\rm{HI}}$ measurement and those from Arecibo and Parkes samples: 
\[ m_{\rm{HI}} = -2.5 \rm{log}(0.2366F_{\rm{HI}}) + 15.84 \]
\[ m_{\rm{HI}- \rm{KLUN}} = m_{\rm{HI}-\rm{Arecibo/Parkes}} - \Delta m_{\rm{HI}} \]
Figure \ref{FigComp} shows the agreement between each telescope's HI magnitude scale after applying this shift.

\begin{figure}
\resizebox{\hsize}{!}{\includegraphics[]{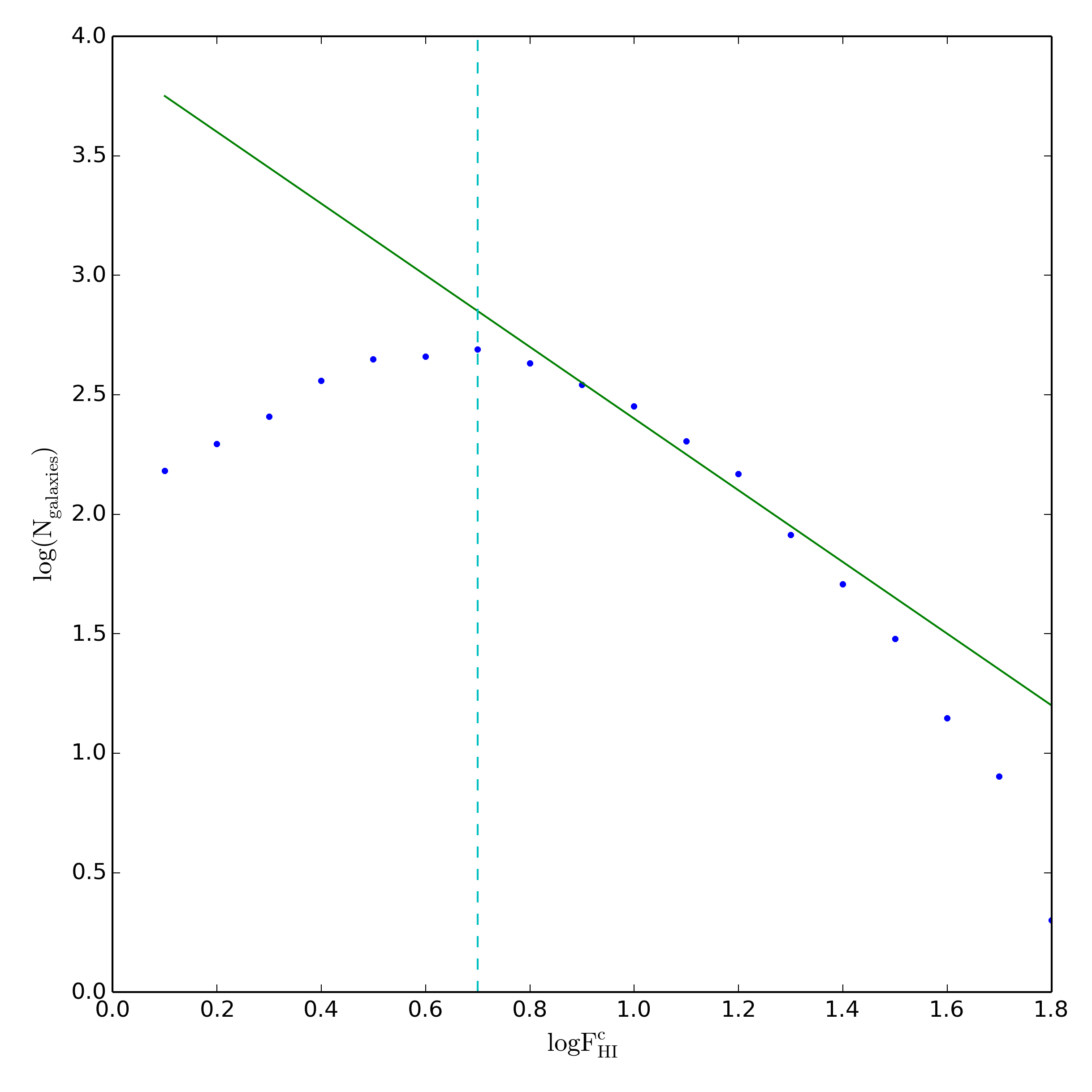}}
\caption{ Logarithmic HI flux distribution.  Blue dots are the galaxy counts per 0.1 dex flux bin; the green line shows the expected growth for a pure flux limited sample. 
The vertical blue dashed line marks the flux limit adopted.} 
\label{FigFlim}
\end{figure}

\begin{figure}
\resizebox{\hsize}{!}{\includegraphics[]{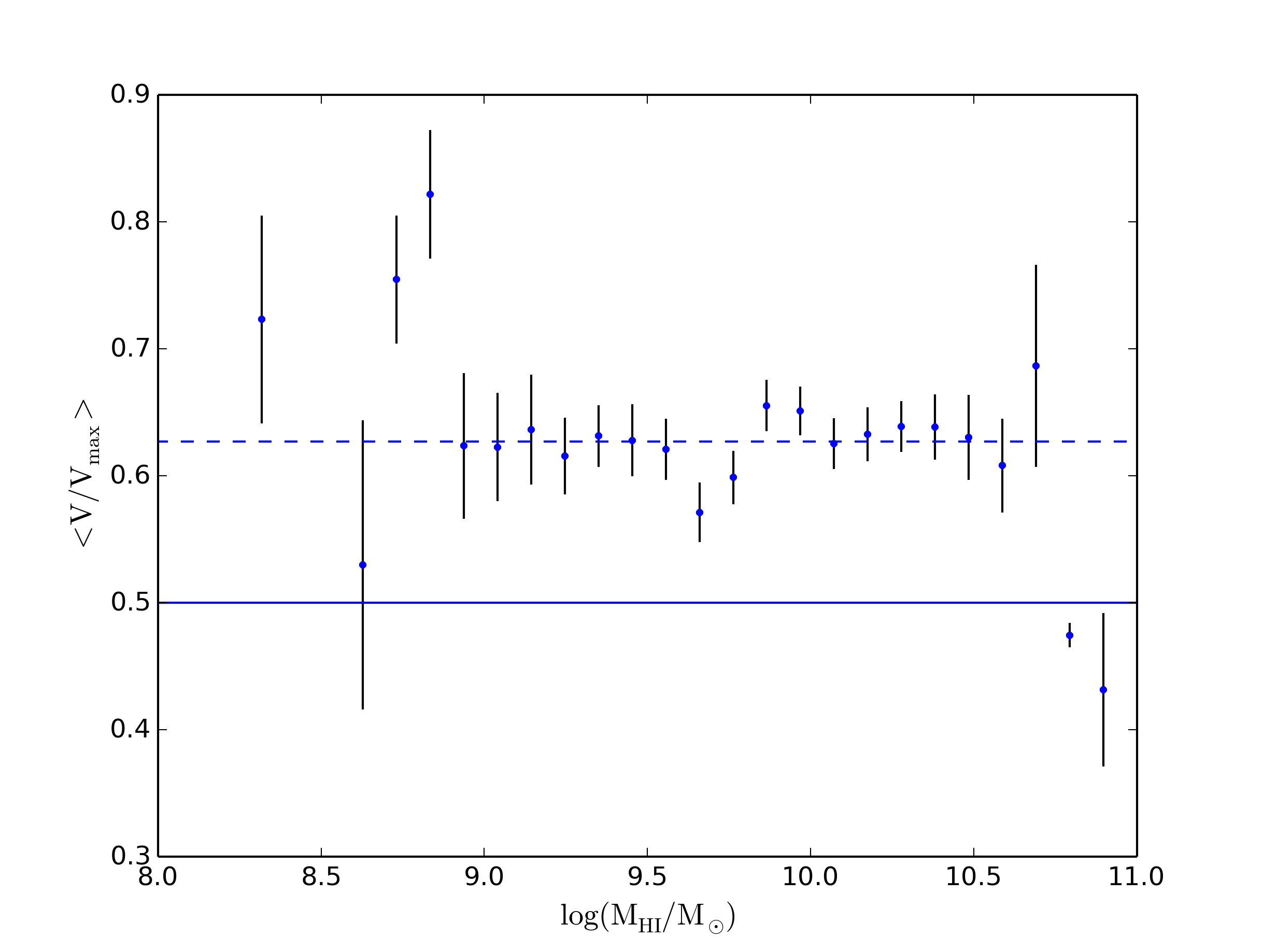}}
\caption{Mean value $< V / V_{max} >$ binned by HI mass. Error bars are Poisson counting uncertainties. 
The solid line indicates the value of 0.5 expected for a homogeneous spatial distribution. 
The dashed line shows the actual average value from the KLUN catalog limited to $\rm{log}(F_{HI}) \geq 0.7$.}
\label{FigVmax}
\end{figure}

\section{ HI mass function from the KLUN compilation}
\label{SecHImass}

\subsection{The $\Sigma \left( 1/ V_{max} \right)$ method}

The HI mass function is generally parametrized as a Schechter function of the form 

\begin{equation}
\phi(M_{\rm HI}) = \frac{d{\rm n}}{d \rm{log} M_{\rm HI}} 
= {\rm ln} 10 \hspace{1.5mm}  \phi_{\star} \left(\frac{M_{\rm HI}}{M_{\odot}} \right) ^{\alpha + 1} {\rm e}^{- \frac{M_{\rm HI}}{M_{\star}}}
,\end{equation}where $\alpha$ represents the faint-end slope, $M_{\star}$ is the characteristic mass, and $\phi_{\star}$ is the scaling or normalization density factor.

To calculate this function we have chosen the simplest way, the so-called $\Sigma \left(1/ V_{max} \right)$ method (\citet{sch68}), which
for a flux limited sample, provides an equivalent of a volume limited catalog. In this method, the source count is weighted by the maximum volume 
$V_{max}$ in which a galaxy of a given HI mass can be detected by the instrument and included in the sample. The value of $\phi(M_{\rm HI})$ in a given 
$M_{\rm HI}$ bin is then obtained by the sum of $1 / V_{max,i}$ for all galaxies in that bin.

The value of $V_{max}$ is calculated as 

\begin{equation}
V_{max}(M_{HI}) = \frac{4 \pi f_{sky}}{3} \left[ \frac{M_{\rm{HI}}}{M_{\odot}} \frac{1}{2.36 \times 10^5 F_{\rm{HI lim}}} \right]^{3/2} \hspace{2.5mm }\rm{Mpc}^3
,\end{equation}where $F_{\rm{HI lim}}$ is the flux limit of the HI survey (assumed complete in flux) and $f_{sky}$ is the sky fraction surveyed.  
Here, by construction, an imprecision in these two  terms directly translates into an imprecision in the normalization density $\phi_{\star}$.

In principle, for a blind survey, $f_{sky}$ should be calculated as the sum of all Nan\c cay beam areas on the sky. 
The Nan\c cay beam size varies with declination so that the half power beam size is 4' $\times$ 22' for $\delta \leq$ 25$^{\circ}$ 
and 4' $\times$ 18'/sin(69.132-$\delta$/2) for $\delta$ > 25$^{\circ}$. This would give us an $f_{sky, t}$ = 2.74 10$^{-3}$ for the whole composite Nan\c cay sample (where ``$t$'' stands  for ``targeted''). 
We know that this number is not the actual $f_{sky}$, since the sample is optically/NIR selected and KLUN galaxy targets have been chosen 
so that no HI measurement was known in the literature at the time of the observation sample definition. However, we can deduce the actual 
$f_{sky}$ from the normalization factor necessary to compare the HI mass function obtained from a targeted survey with the value obtained from a pure blind search catalog. 
This is done by forcing the KLUN HI mass Schechter function to have the same $\log{(\phi_*)}$ as that from the ALFALFA catalog, which is used here as the reference blind survey.
From this $\log{(\phi_*)}$ comparison,  we get a factor of 112 on $f_{sky}$, which thus becomes $f_{sky}$ = $f_{sky, t} \times 112$ = 0.307
for the actual value to be used in Eq. (2).

\begin{figure*}
\centering
\includegraphics[width=16.2cm]{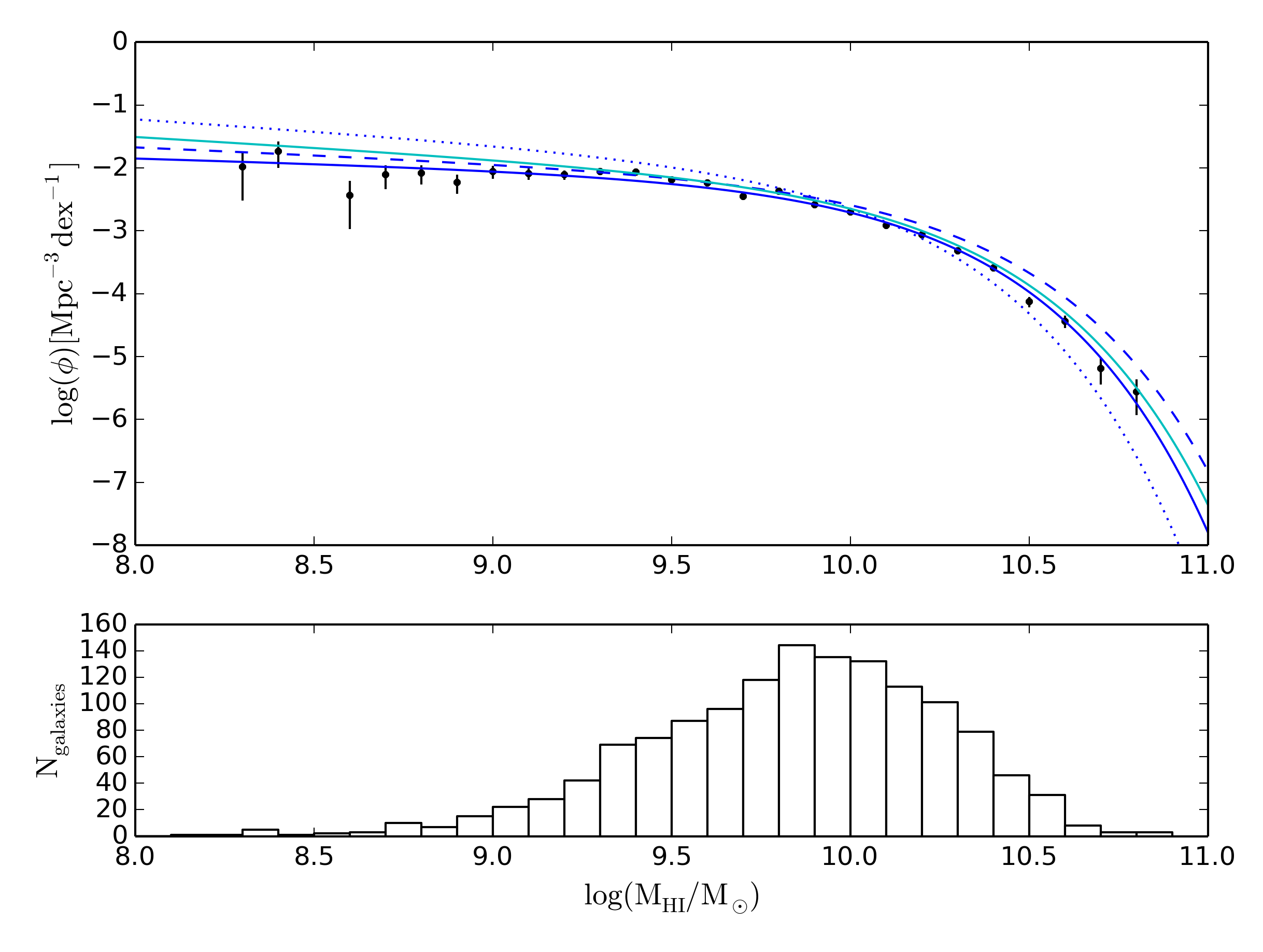}
\caption{HI mass function from KLUN (dark blue line) vs. \citet{spr05} (blue dashed line),  \citet{zwa05} (blue dotted line), and \citet{hay11} (cyan line).  KLUN and Springob results are both normalized to the ALFALFA survey.
Observed values (black dots) are calculated per bins of 0.1 in $\log(M_{HI}/M_{\odot})$ with error bars from the Poisson law, and the corresponding population is shown as a histogram in the bottom panel.
} 
\label{FigHImass}
\end{figure*}

Using the $\Sigma \left(1/ V_{max} \right)$ method requires two main assumptions: 1) we control the 
selection of the catalog, and 2) the space distribution of the sources is not too far from homogeneity. To deal with the first point, we forced a sharp cutoff 
to the KLUN Nan\c cay HI-catalog at the apparent completeness limit $\rm{log}(F_{HI}) \geq 0.7$ (see Fig. \ref{FigFlim}). We assume here that this limit in HI flux
is stronger than the optical or NIR selection limit of the parent sample used to define the radio follow-up program. We note that this is not necessarily 
true for the very low surface brightness population. Concerning the second point, the low spatial resolution of single-dish observations at 1.4 GHz 
leads to HI line confusion at the scale of the antenna beam ($\equiv 4' \times 22'$) and limits high spatial clustering of HI catalogs. Fig. \ref{FigStat} shows the 
kinematical distance distribution of the sample. The distribution is rather smoothed with a few peaks emerging at 2\,500 km.s$^{-1}$, 
3\,800-5\,000 km.s$^{-1}$, 5\,700 km.s$^{-1}$, and 6\,800 km.s$^{-1}$, which correspond to the main supercluster concentrations known in the area 
visible from Nan\c cay (e.g., respectively Hydra and Antlia clusters, Centaurus and Perseus-Pisces complexes, Abell347 cluster, and Coma cluster).
A method for controlling the deviation from a pure homogeneous spatial distribution is known as the $V / V_{max} $ statistical test (\citet{sch68}),
which compares the actual volume within the source distance to the maximum volume in which the source might be detected. In a homogeneous volume,
the expectation value $< V / V_{max} >$ is 0.5. Figure \ref{FigVmax} shows the $V / V_{max} $ per bin of $M_{\rm HI}$. 
The values are consistent with  $< V / V_{max} > \equiv 0.62$ indicating the presence of overdensities, as expected from the various peaks appearing 
in the radial velocity distribution (Fig. \ref{FigStat}). This suggests  that the sample might be biased by the few large clusters present in this redshift domain.

Finally, as discussed in \citet{mar10}, in the local volume where peculiar velocities are comparable to the expansion, 
the distance uncertainty introduced by the use of a pure Hubble flow for the distance estimate may cause distortion of 
the faint end of the HI mass function. The infall towards the Virgo cluster being the most important feature at this scale, 
we then used radial velocities corrected for a Virgo infall model, $V_{Vir}$, for the kinematical distance estimate. 
According to hyperleda \footnote{http://leda.univ-lyon1.fr/leda/param/vvir.html}, $V_{Vir} = V_{LG} + 208 \hspace{1.5mm} \rm{cos}(\theta)$,  
where $V_{LG}$ is the velocity relative to the Local Group, 208 km.s$^{-1}$ is the infall velocity of the Local Group according to \citet{the98b} and  \citet{ter02}, 
and  $\theta$ is the angular distance between the observed direction and the direction of the center of the Virgo cluster (SG 104,-2 or J123310+112152).

\subsection{Results}
\label{SecKLUNres}

We show in Fig. \ref{FigHImass} the HI mass function calculated for the whole flux limited sample, and the corresponding Schechter function together 
with three previous results from \citet{spr05}, \citet{zwa05}, and \citet{mar10}. 
The HI mass scale has been homogenized between the different references to take into account the shift measured between 
Arecibo, Nan\c cay, and Parkes HI fluxes discussed in paragraph~\ref{HIflu}.

For the KLUN sample, we obtain the following parameter values (after normalization on $f_{sky}$ according to the previous section)
\[ \alpha = -1.16, \hspace{2.5mm}
 \rm{log} \left( \frac{M_{\star}}{M_{\odot}} \right) = 9.90 , \hspace{2.5mm}
 \phi_{\star} = 3.12 \hspace{1.0mm} 10^{-3} \hspace{1.5mm} Mpc^{-3}, \]

In Fig. \ref{FigHImass}, the KLUN and \citet{spr05} HI mass functions have been normalized to the same $\phi_{\star}$ 
derived for ALFALFA  \citet{mar10}.  For these two catalogs, which are based on targeted observations and not on a blind HI search, 
the sky filling factor is not directly available or comparable owing to the selection and identification process based on optical or NIR selection (see previous section). 
The result from HIPASS, which is also a blind HI survey, is shown without any normalization.

\section{Conclusions}

We have presented here the last HI data release of a long series of observations with the NRT in the context of Tully-Fisher applications. 
A large fraction of these remaining 828 galaxies of the program is   a compilation of low S/N or problematic cases;  
however, this subset contains 417 ($\sim$ 50\%) of good and high quality HI line profile 
measurements (classes A and B). Together with the  previous publications from the same observational program (KLUN), we have  
produced -- as a by-product of the original Tully-Fisher program -- a homogeneous sample of HI-fluxes for 4876 spiral galaxies, 
which constitutes a useful catalog for HI mass functions studies. This catalog is complete down to an HI flux of 5 Jy km.s$^{-1}$ 
and  samples the radial velocity space between 2000 and 8000 km.s$^{-1}$.Mpc$^{-1}$. We  finally obtain a first estimation of the Schechter  
parameters for the HI mass function in the range from 5 10$^8$ to 5 10$^{10}$ solar masses, and we obtain a solution 
which is perfectly coherent with previous studies.

\begin{acknowledgements} 
The Nan{\c c}ay radio Observatory is operated by the Paris Observatory, associated with the French Centre National de la Recherche Scientifique (CNRS)
and with the University of Orl\'eans.
\end{acknowledgements}


\bibliographystyle{aa}
\bibliography{MS29813_final}

\begin{thebibliography}{20}
\expandafter\ifx\csname natexlab\endcsname\relax\def\natexlab#1{#1}\fi

\bibitem[{{Bottinelli} {et~al.}(1993){Bottinelli}, {Durand}, {Fouque},
  {Garnier}, {Gouguenheim}, {Loulergue}, {Paturel}, {Petit}, \&
  {Teerikorpi}}]{bot93}
{Bottinelli}, L., {Durand}, N., {Fouque}, P., {et~al.} 1993, \aaps, 102, 57

\bibitem[{{Bottinelli} {et~al.}(1992){Bottinelli}, {Durand}, {Fouque},
  {Garnier}, {Gouguenheim}, {Paturel}, \& {Teerikorpi}}]{bot92}
{Bottinelli}, L., {Durand}, N., {Fouque}, P., {et~al.} 1992, \aaps, 93, 173

\bibitem[{{Bottinelli} {et~al.}(1990){Bottinelli}, {Gouguenheim}, {Fouque}, \&
  {Paturel}}]{bot90}
{Bottinelli}, L., {Gouguenheim}, L., {Fouque}, P., \& {Paturel}, G. 1990,
  \aaps, 82, 391

\bibitem[{{di Nella} {et~al.}(1996){di Nella}, {Paturel}, {Walsh},
  {Bottinelli}, {Gouguenheim}, \& {Theureau}}]{nel96}
{di Nella}, H., {Paturel}, G., {Walsh}, A.~J., {et~al.} 1996, \aaps, 118, 311

\bibitem[{{Fouque} {et~al.}(1990){Fouque}, {Bottinelli}, {Gouguenheim}, \&
  {Paturel}}]{fou90}
{Fouque}, P., {Bottinelli}, L., {Gouguenheim}, L., \& {Paturel}, G. 1990, \apj,
  349, 1

\bibitem[{{Haynes} {et~al.}(2011){Haynes}, {Giovanelli}, {Martin}, {Hess},
  {Saintonge}, {Adams}, {Hallenbeck}, {Hoffman}, {Huang}, {Kent}, {Koopmann},
  {Papastergis}, {Stierwalt}, {Balonek}, {Craig}, {Higdon}, {Kornreich},
  {Miller}, {O'Donoghue}, {Olowin}, {Rosenberg}, {Spekkens}, {Troischt}, \&
  {Wilcots}}]{hay11}
{Haynes}, M.~P., {Giovanelli}, R., {Martin}, A.~M., {et~al.} 2011, \aj, 142,
  170

\bibitem[{{Jarrett} {et~al.}(2000){Jarrett}, {Chester}, {Cutri}, {Schneider},
  {Rosenberg}, {Huchra}, \& {Mader}}]{jar00}
{Jarrett}, T.-H., {Chester}, T., {Cutri}, R., {et~al.} 2000, \aj, 120, 298

\bibitem[{{Martin} {et~al.}(2010){Martin}, {Papastergis}, {Giovanelli},
  {Haynes}, {Springob}, \& {Stierwalt}}]{mar10}
{Martin}, A.~M., {Papastergis}, E., {Giovanelli}, R., {et~al.} 2010, \apj, 723,
  1359

\bibitem[{{Meyer} {et~al.}(2004){Meyer}, {Zwaan}, {Webster}, {Staveley-Smith},
  {Ryan-Weber}, {Drinkwater}, {Barnes}, {Howlett}, {Kilborn}, {Stevens},
  {Waugh}, {Pierce}, {Bhathal}, {de Blok}, {Disney}, {Ekers}, {Freeman},
  {Garcia}, {Gibson}, {Harnett}, {Henning}, {Jerjen}, {Kesteven}, {Knezek},
  {Koribalski}, {Mader}, {Marquarding}, {Minchin}, {O'Brien}, {Oosterloo},
  {Price}, {Putman}, {Ryder}, {Sadler}, {Stewart}, {Stootman}, \&
  {Wright}}]{mey04}
{Meyer}, M.~J., {Zwaan}, M.~A., {Webster}, R.~L., {et~al.} 2004, \mnras, 350,
  1195

\bibitem[{{Paturel} {et~al.}(2003{\natexlab{a}}){Paturel}, {Petit}, {Prugniel},
  {Theureau}, {Rousseau}, {Brouty}, {Dubois}, \& {Cambr{\'e}sy}}]{pat03a}
{Paturel}, G., {Petit}, C., {Prugniel}, P., {et~al.} 2003{\natexlab{a}}, \aap,
  412, 45

\bibitem[{{Paturel} {et~al.}(2003{\natexlab{b}}){Paturel}, {Theureau},
  {Bottinelli}, {Gouguenheim}, {Coudreau-Durand}, {Hallet}, \&
  {Petit}}]{pat03b}
{Paturel}, G., {Theureau}, G., {Bottinelli}, L., {et~al.} 2003{\natexlab{b}},
  \aap, 412, 57

\bibitem[{{Paturel} {et~al.}(2005){Paturel}, {Vauglin}, {Petit},
  {Borsenberger}, {Epchtein}, {Fouqu{\'e}}, \& {Mamon}}]{pat05}
{Paturel}, G., {Vauglin}, I., {Petit}, C., {et~al.} 2005, \aap, 430, 751

\bibitem[{{Schmidt}(1968)}]{sch68}
{Schmidt}, M. 1968, \apj, 151, 393

\bibitem[{{Springob} {et~al.}(2005){Springob}, {Haynes}, {Giovanelli}, \&
  {Kent}}]{spr05}
{Springob}, C.~M., {Haynes}, M.~P., {Giovanelli}, R., \& {Kent}, B.~R. 2005,
  \apjs, 160, 149

\bibitem[{{Terry} {et~al.}(2002){Terry}, {Paturel}, \& {Ekholm}}]{ter02}
{Terry}, J.~N., {Paturel}, G., \& {Ekholm}, T. 2002, \aap, 393, 57

\bibitem[{{Theureau} {et~al.}(1998{\natexlab{a}}){Theureau}, {Bottinelli},
  {Coudreau-Durand}, {Gouguenheim}, {Hallet}, {Loulergue}, {Paturel}, \&
  {Teerikorpi}}]{the98a}
{Theureau}, G., {Bottinelli}, L., {Coudreau-Durand}, N., {et~al.}
  1998{\natexlab{a}}, \aaps, 130, 333

\bibitem[{{Theureau} {et~al.}(2005){Theureau}, {Coudreau}, {Hallet}, {Hanski},
  {Alsac}, {Bottinelli}, {Gouguenheim}, {Martin}, \& {Paturel}}]{the05}
{Theureau}, G., {Coudreau}, N., {Hallet}, N., {et~al.} 2005, \aap, 430, 373

\bibitem[{{Theureau} {et~al.}(2007){Theureau}, {Hanski}, {Coudreau}, {Hallet},
  \& {Martin}}]{the07}
{Theureau}, G., {Hanski}, M.~O., {Coudreau}, N., {Hallet}, N., \& {Martin},
  J.-M. 2007, \aap, 465, 71

\bibitem[{{Theureau} {et~al.}(1998{\natexlab{b}}){Theureau}, {Rauzy},
  {Bottinelli}, \& {Gouguenheim}}]{the98b}
{Theureau}, G., {Rauzy}, S., {Bottinelli}, L., \& {Gouguenheim}, L.
  1998{\natexlab{b}}, \aap, 340, 21

\bibitem[{{Zwaan} {et~al.}(2005){Zwaan}, {Meyer}, {Staveley-Smith}, \&
  {Webster}}]{zwa05}
{Zwaan}, M.~A., {Meyer}, M.~J., {Staveley-Smith}, L., \& {Webster}, R.~L. 2005,
  \mnras, 359, L30

\end{thebibliography}


\begin{appendix}
\section{HI profiles. Continuation of Figure 2}

\begin{figure*}[ht]
\centering
\includegraphics[width=\textwidth]{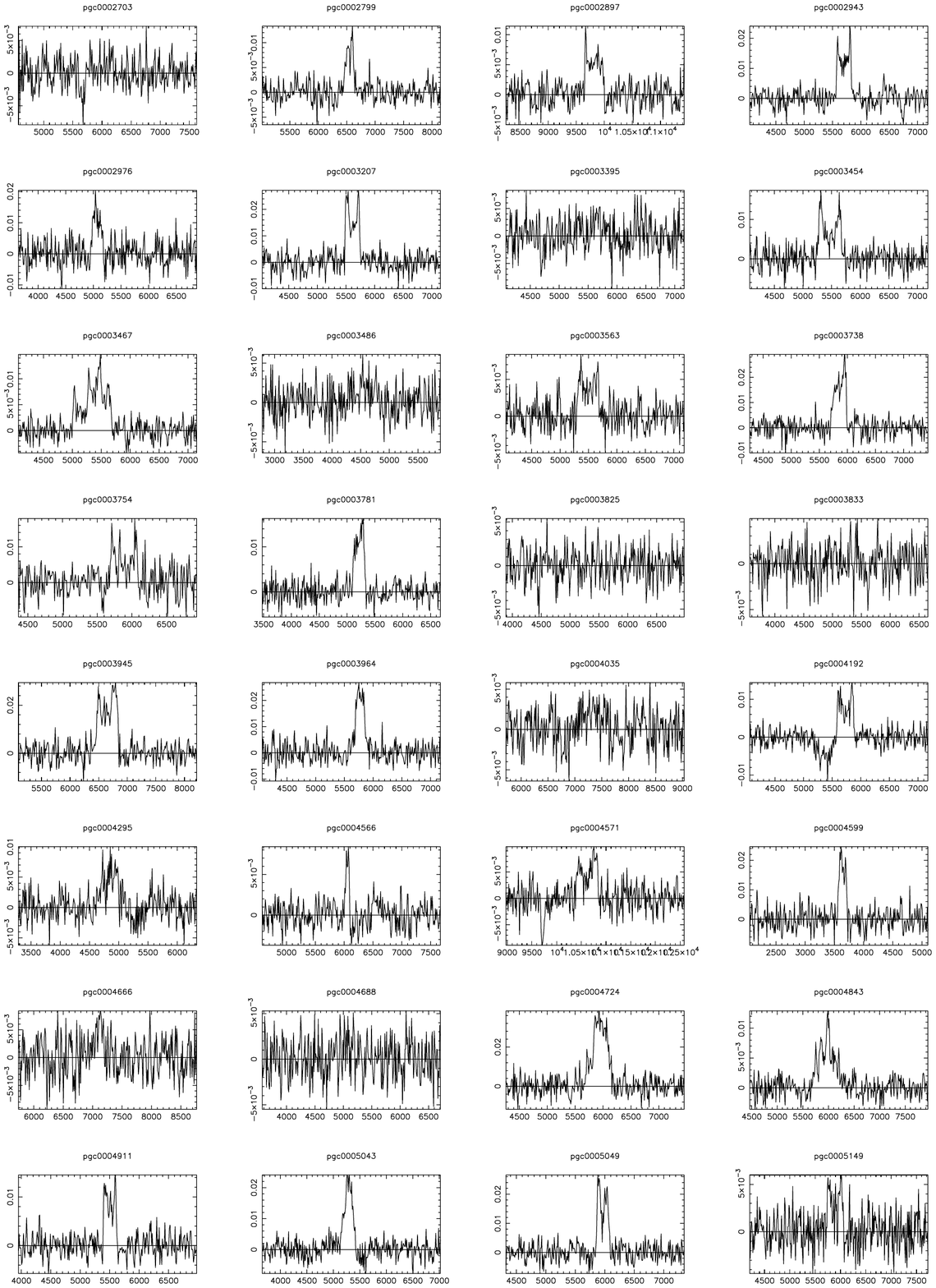}
\captcont{\small Fig. 2 {\bf b.} HI profiles. Continuation of Fig2.}
\end{figure*}

\clearpage
\newpage

\begin{figure*}
\centering
\includegraphics[width=\textwidth]{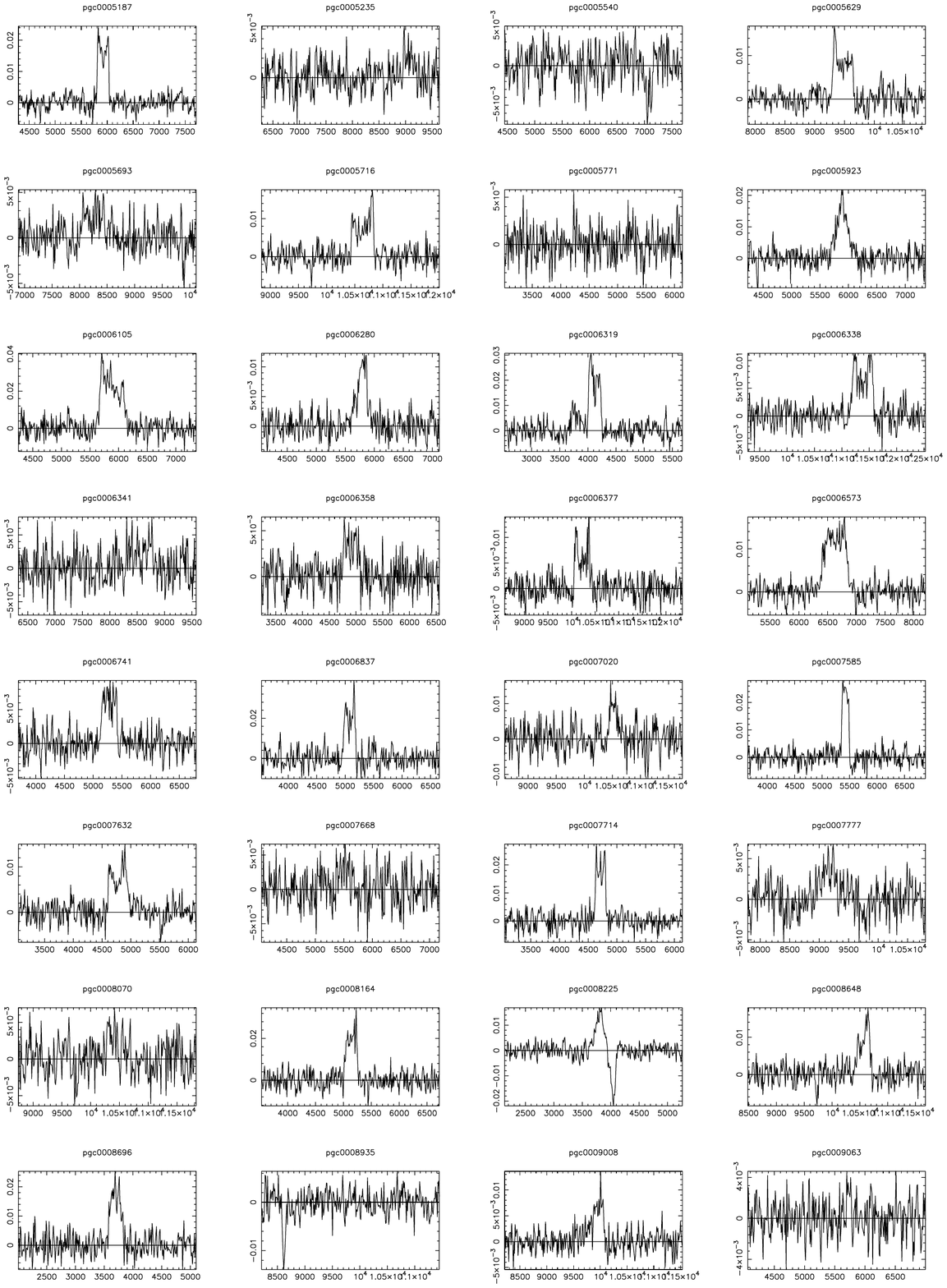}
\captcont{\small Fig. 2  {\bf c.} HI profiles. Continued.}
\end{figure*}

\clearpage
\newpage

\begin{figure*}
\centering
\includegraphics[width=\textwidth]{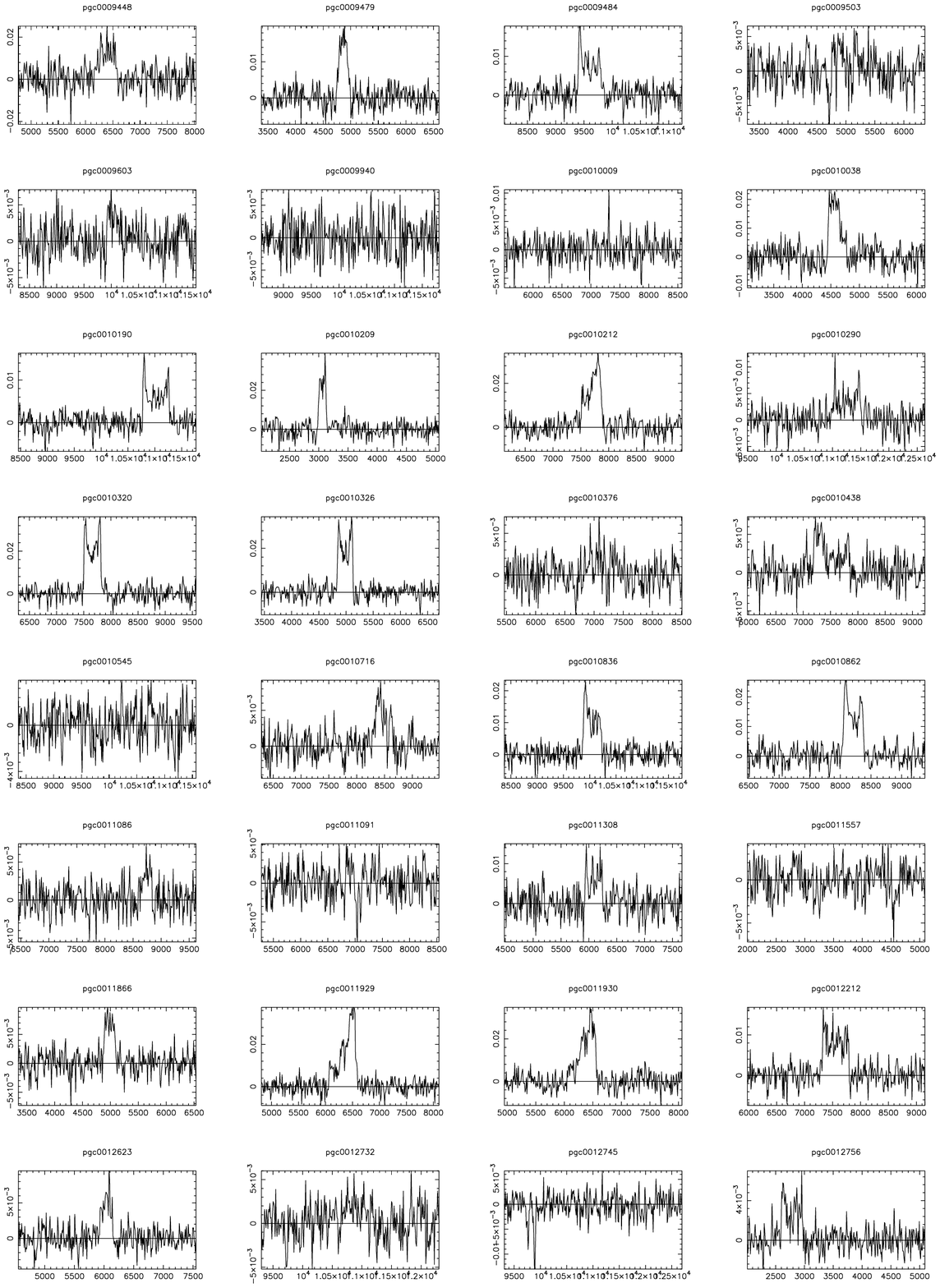}
\captcont{\small Fig. 2 {\bf d.} HI profiles. Continued.}
\end{figure*}

\clearpage
\newpage

\begin{figure*}
\centering
\includegraphics[width=\textwidth]{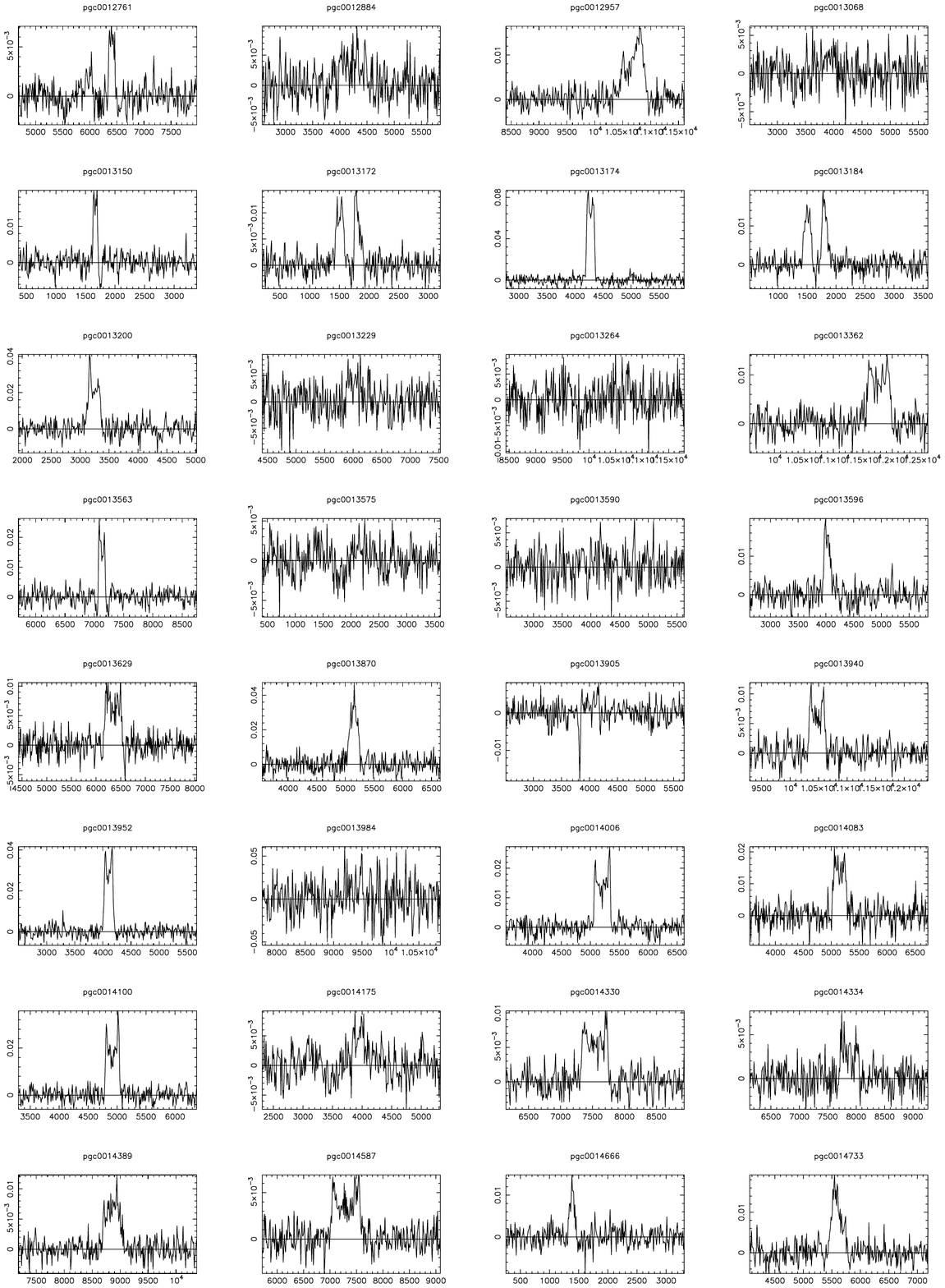}
\captcont{\small Fig. 2  {\bf e.} HI profiles. Continued.}
\end{figure*}

\clearpage
\newpage

\begin{figure*}
\centering
\includegraphics[width=\textwidth]{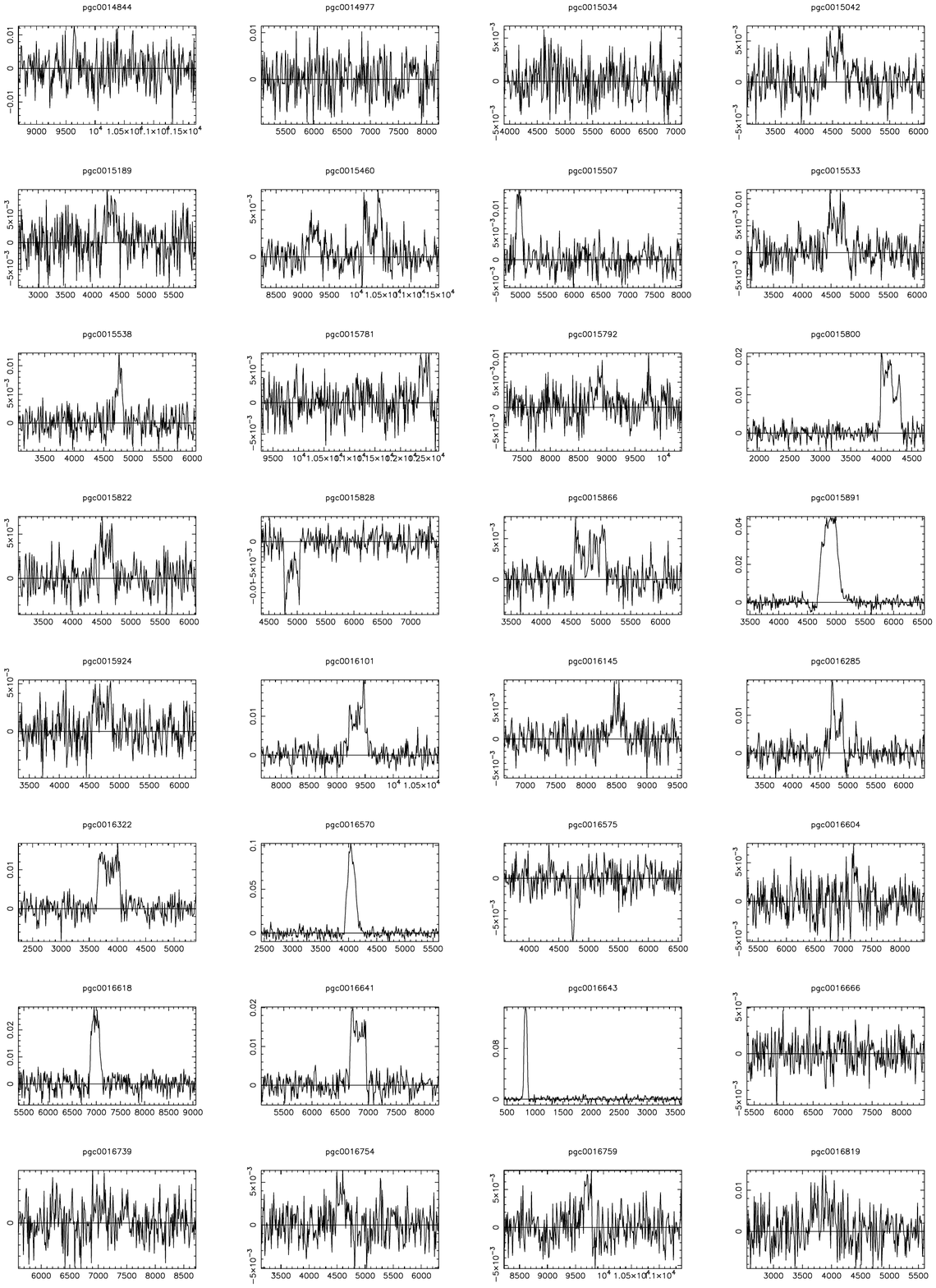}
\captcont{\small Fig. 2 {\bf f.} HI profiles. Continued.}
\end{figure*}

\clearpage
\newpage

\begin{figure*}
\centering
\includegraphics[width=\textwidth]{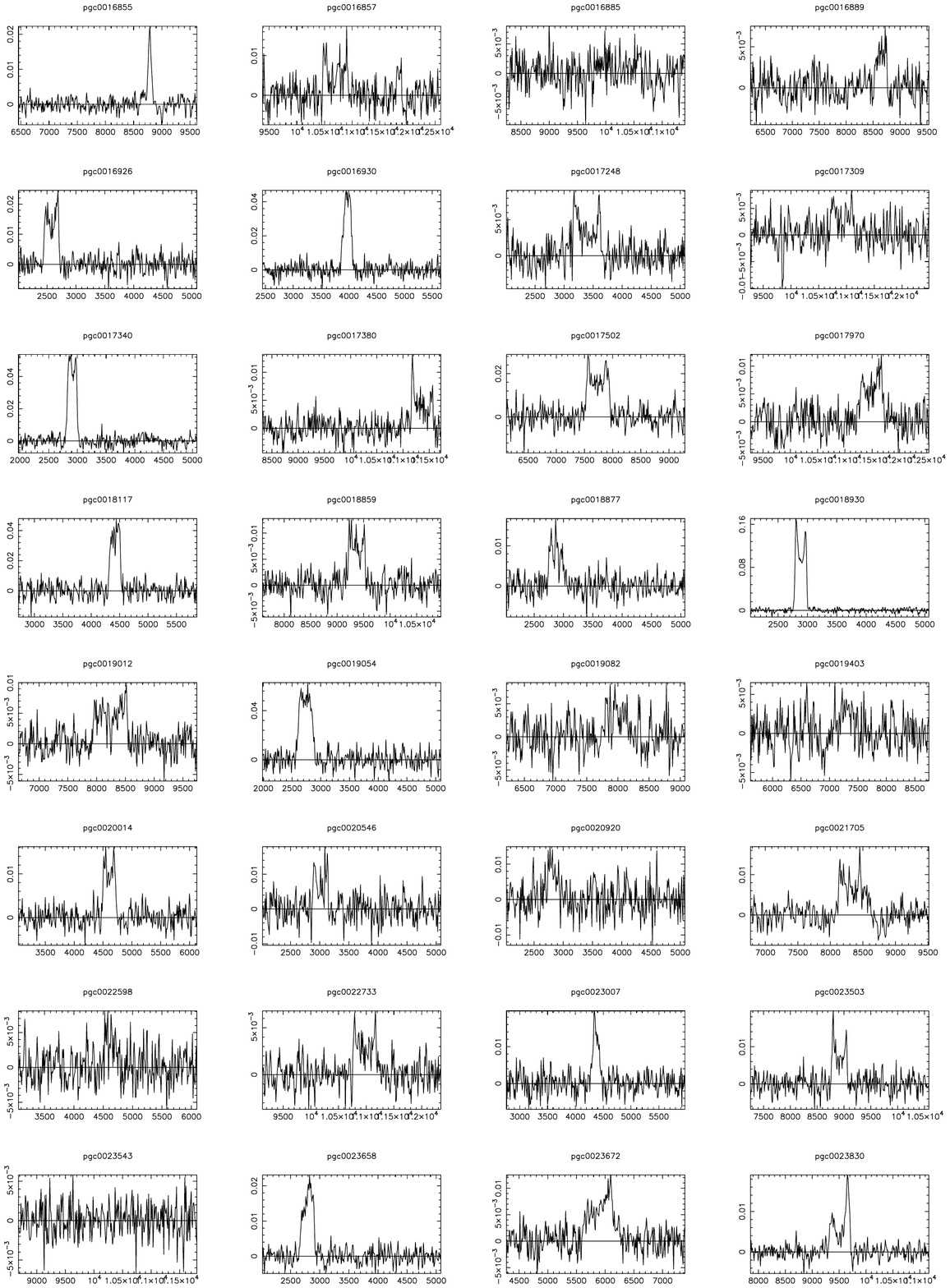}
\captcont{\small Fig. 2  {\bf g.} HI profiles. Continued.}
\end{figure*}

\clearpage
\newpage

\begin{figure*}
\centering
\includegraphics[width=\textwidth]{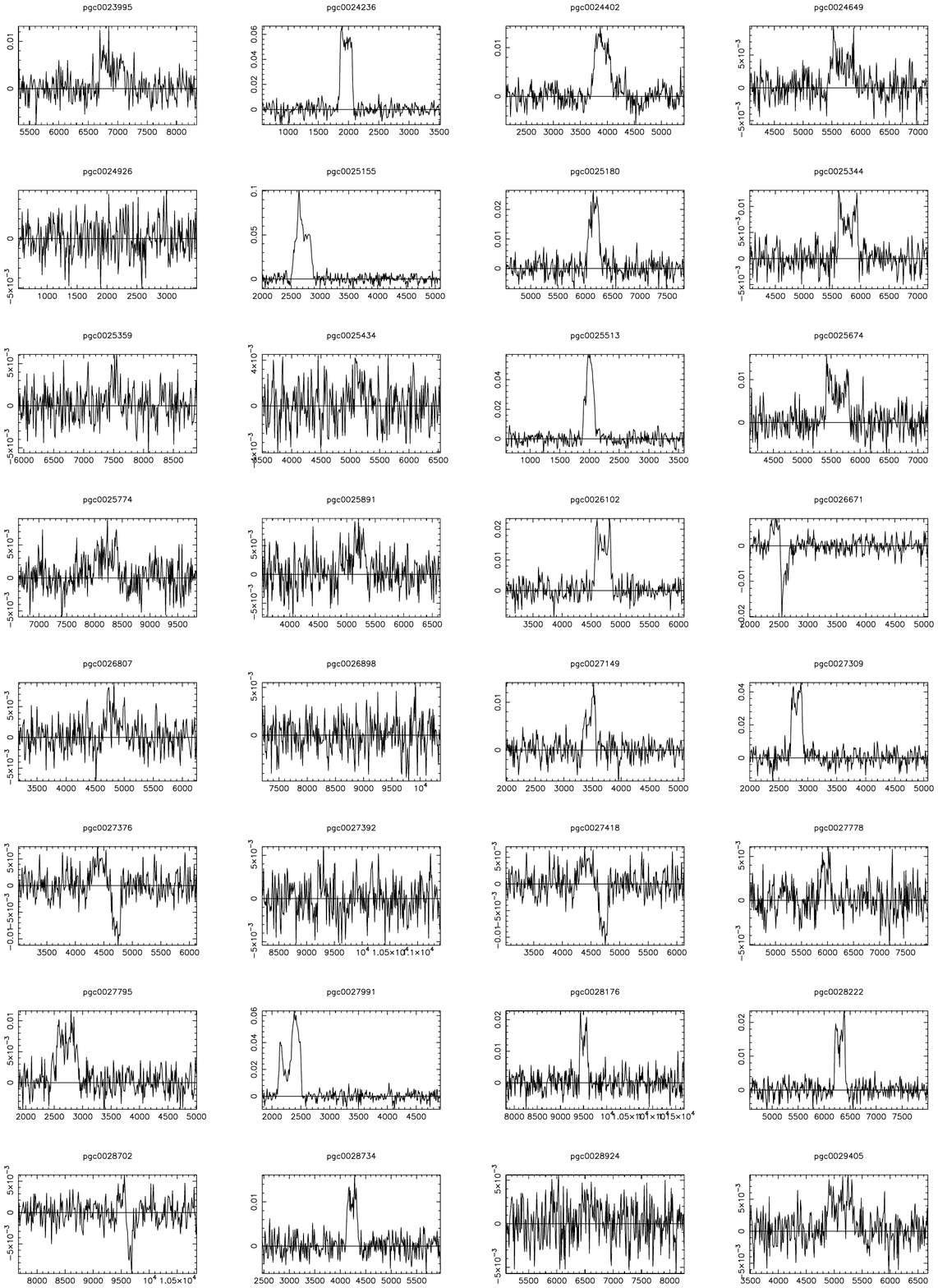}
\captcont{\small Fig. 2  {\bf h.} HI profiles. Continued.}
\end{figure*}

\clearpage
\newpage

\begin{figure*}
\centering
\includegraphics[width=\textwidth]{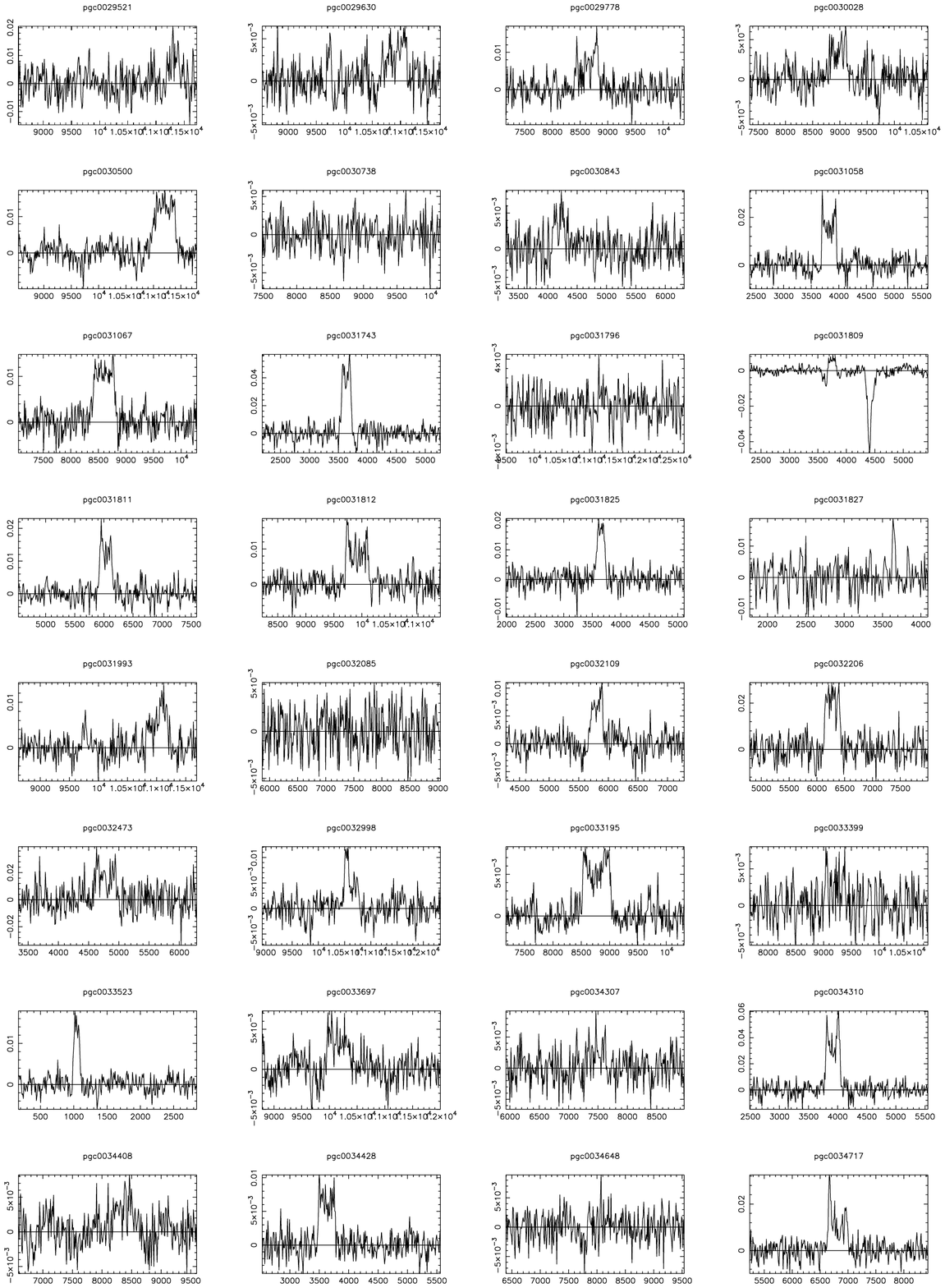}
\captcont{\small Fig. 2  {\bf i.} HI profiles. Continued.}
\end{figure*}

\clearpage
\newpage

\begin{figure*}
\centering
\includegraphics[width=\textwidth]{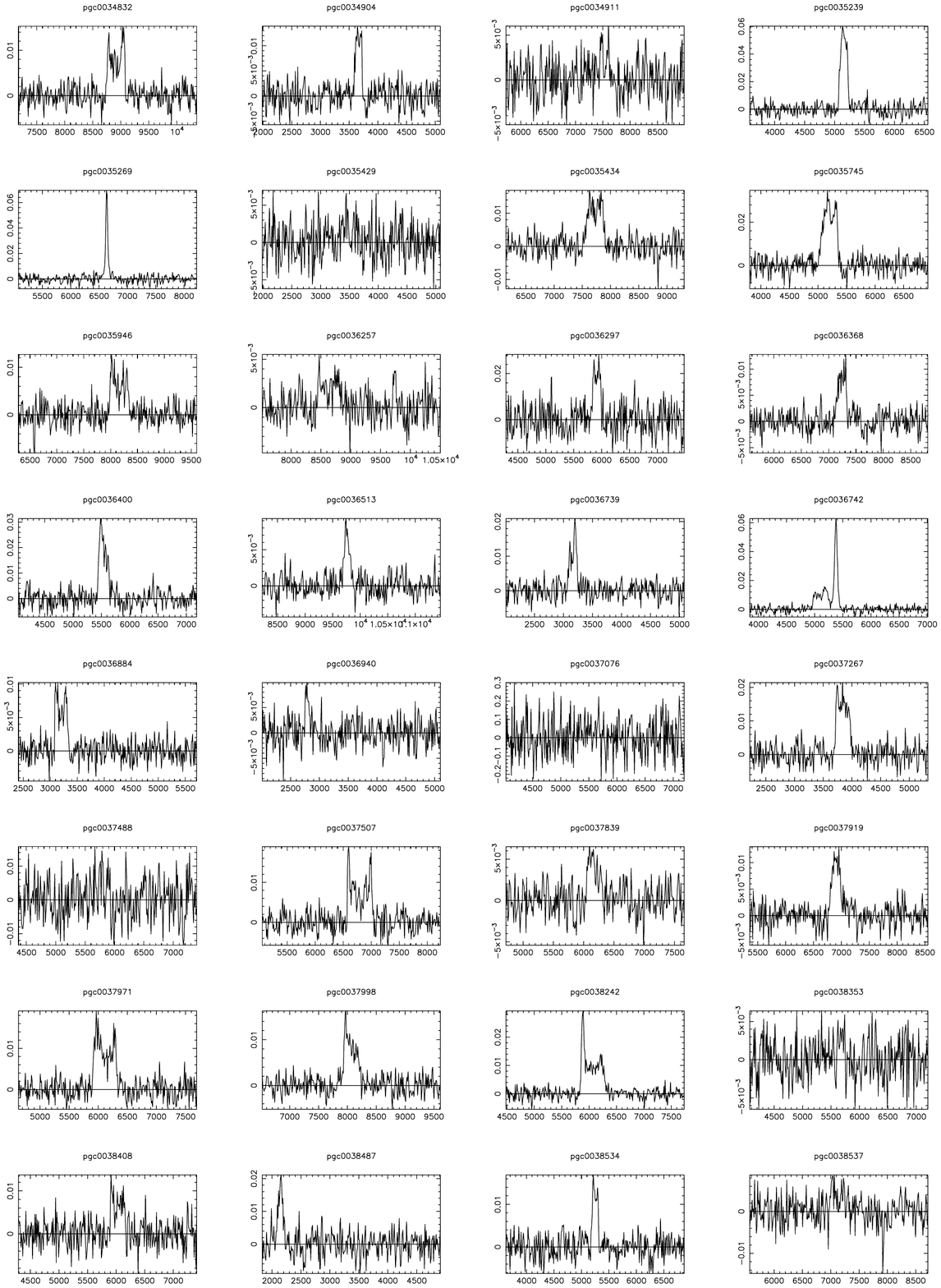}
\captcont{\small Fig. 2 {\bf j.} HI profiles. Continued.}
\end{figure*}

\clearpage
\newpage

\begin{figure*}
\centering
\includegraphics[width=\textwidth]{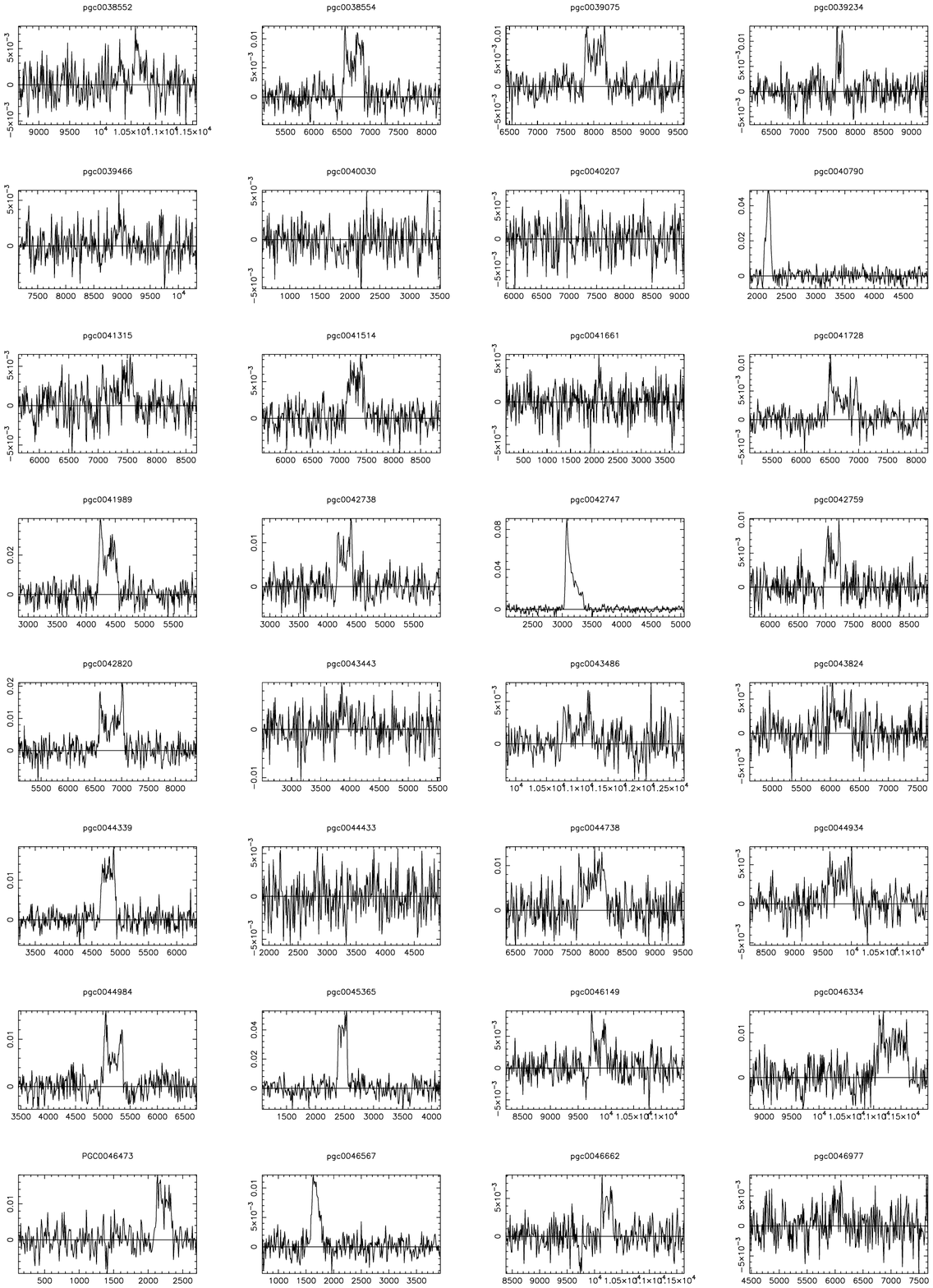}
\captcont{\small Fig. 2 {\bf k.} HI profiles. Continued.}
\end{figure*}

\newpage

\begin{figure*}
\centering
\includegraphics[width=\textwidth]{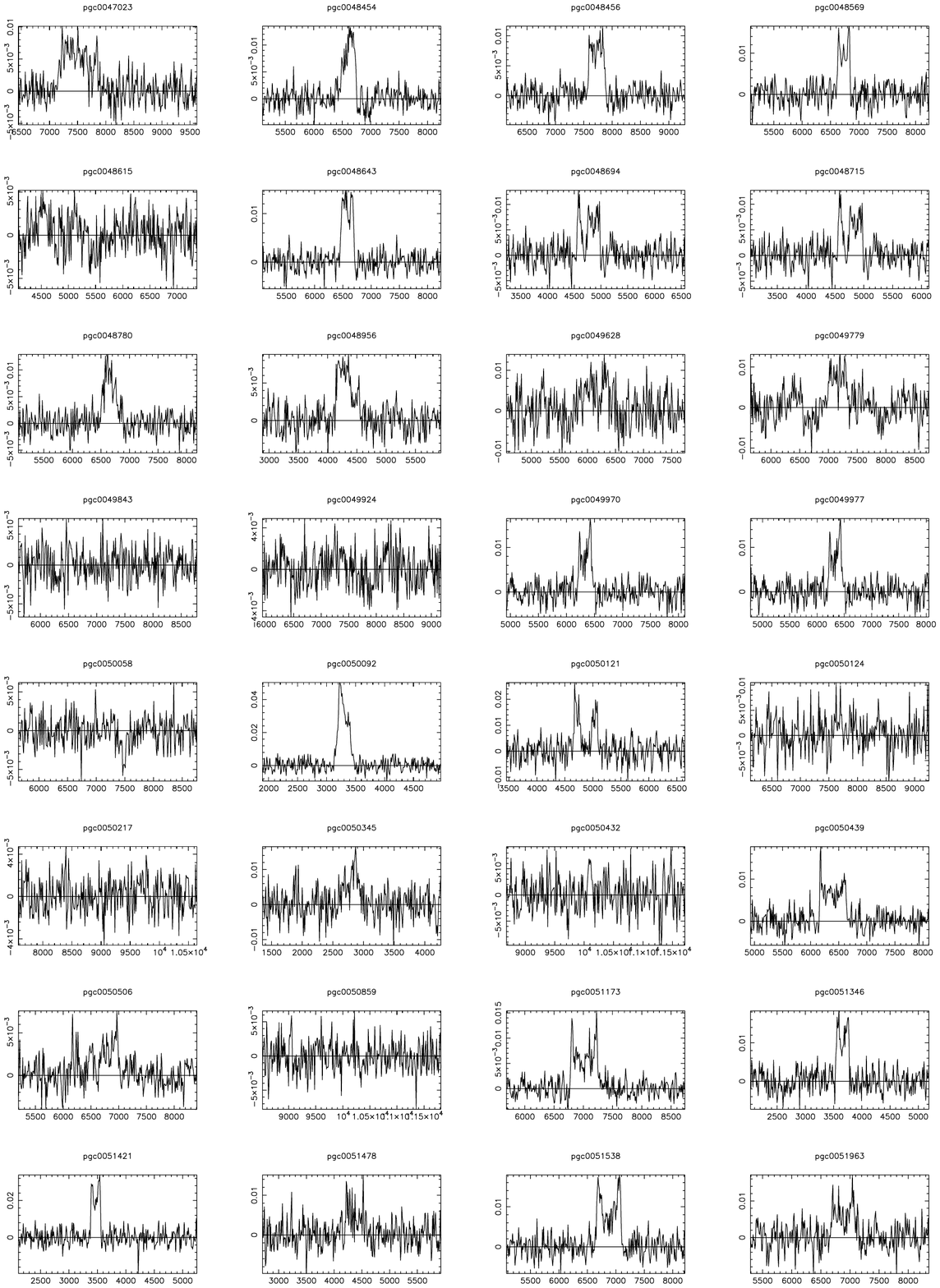}
\captcont{\small Fig. 2  {\bf l.} HI profiles. Continued.}
\end{figure*}

\clearpage
\newpage

\begin{figure*}
\centering
\includegraphics[width=\textwidth]{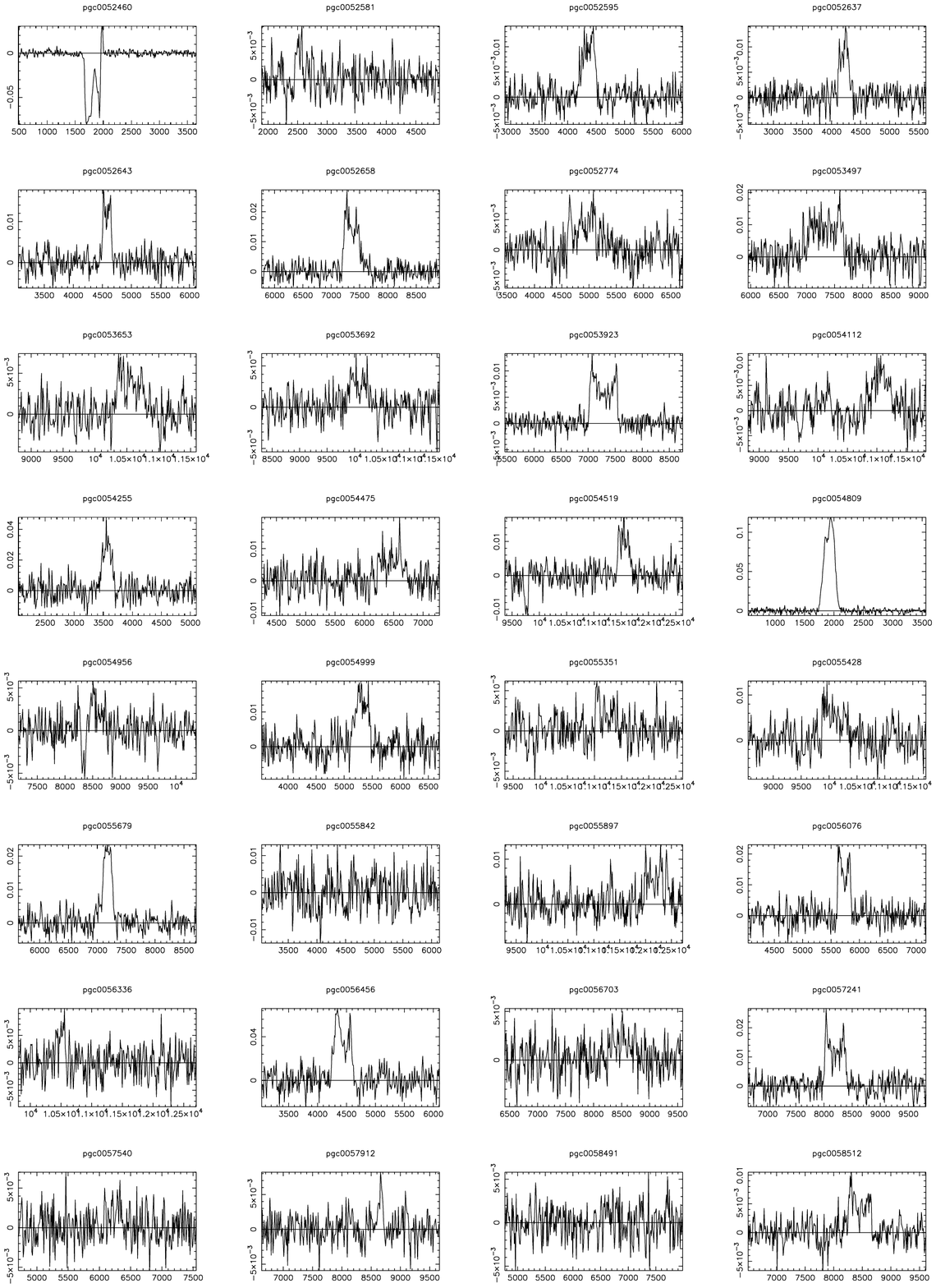}
\captcont{\small Fig. 2  {\bf m.} HI profiles. Continued.}
\end{figure*}

\clearpage
\newpage

\begin{figure*}
\centering
\includegraphics[width=\textwidth]{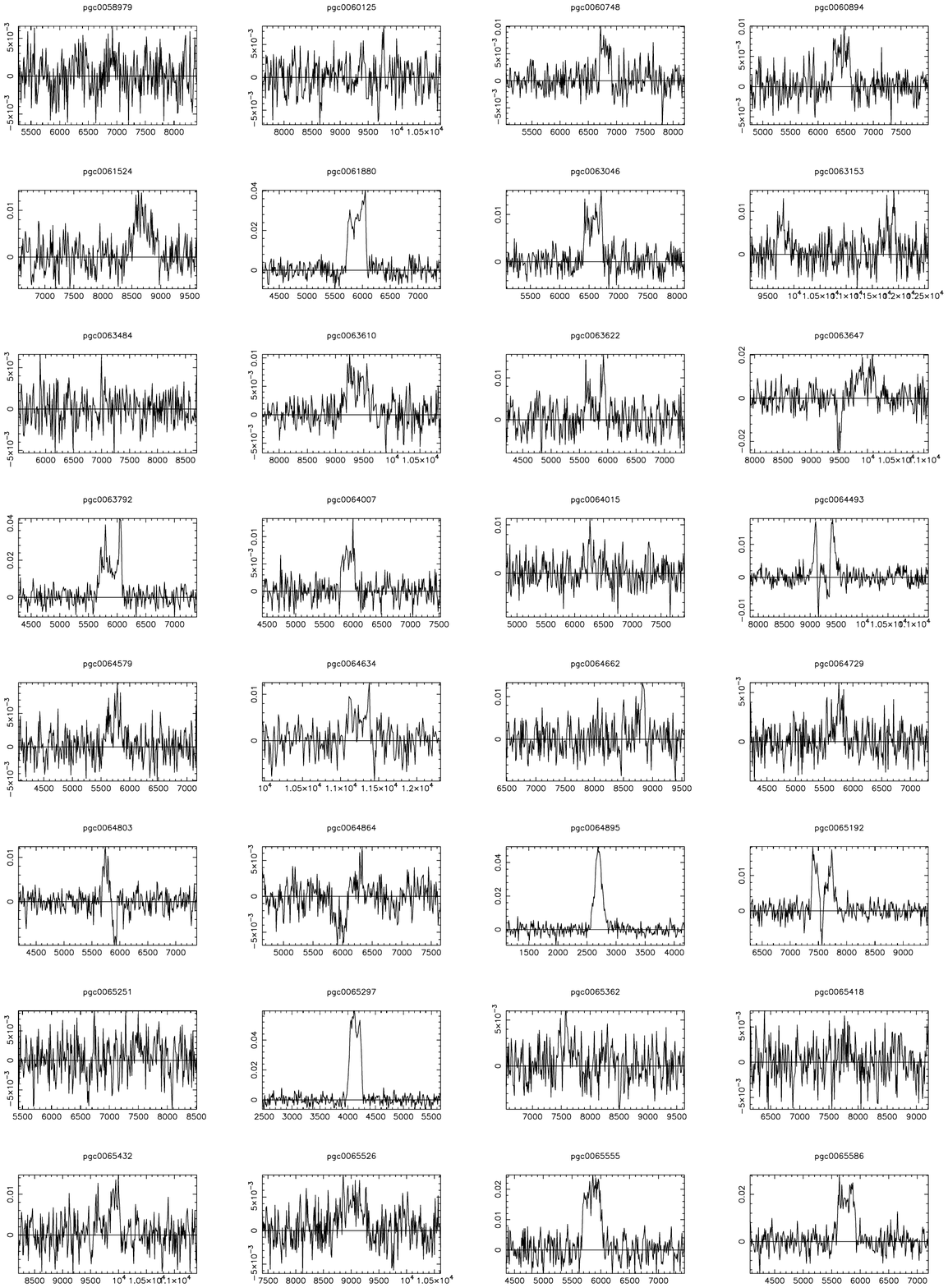}
\captcont{\small Fig. 2 {\bf n.} HI profiles. Continued.}
\end{figure*}

\clearpage
\newpage

\begin{figure*}
\centering
\includegraphics[width=\textwidth]{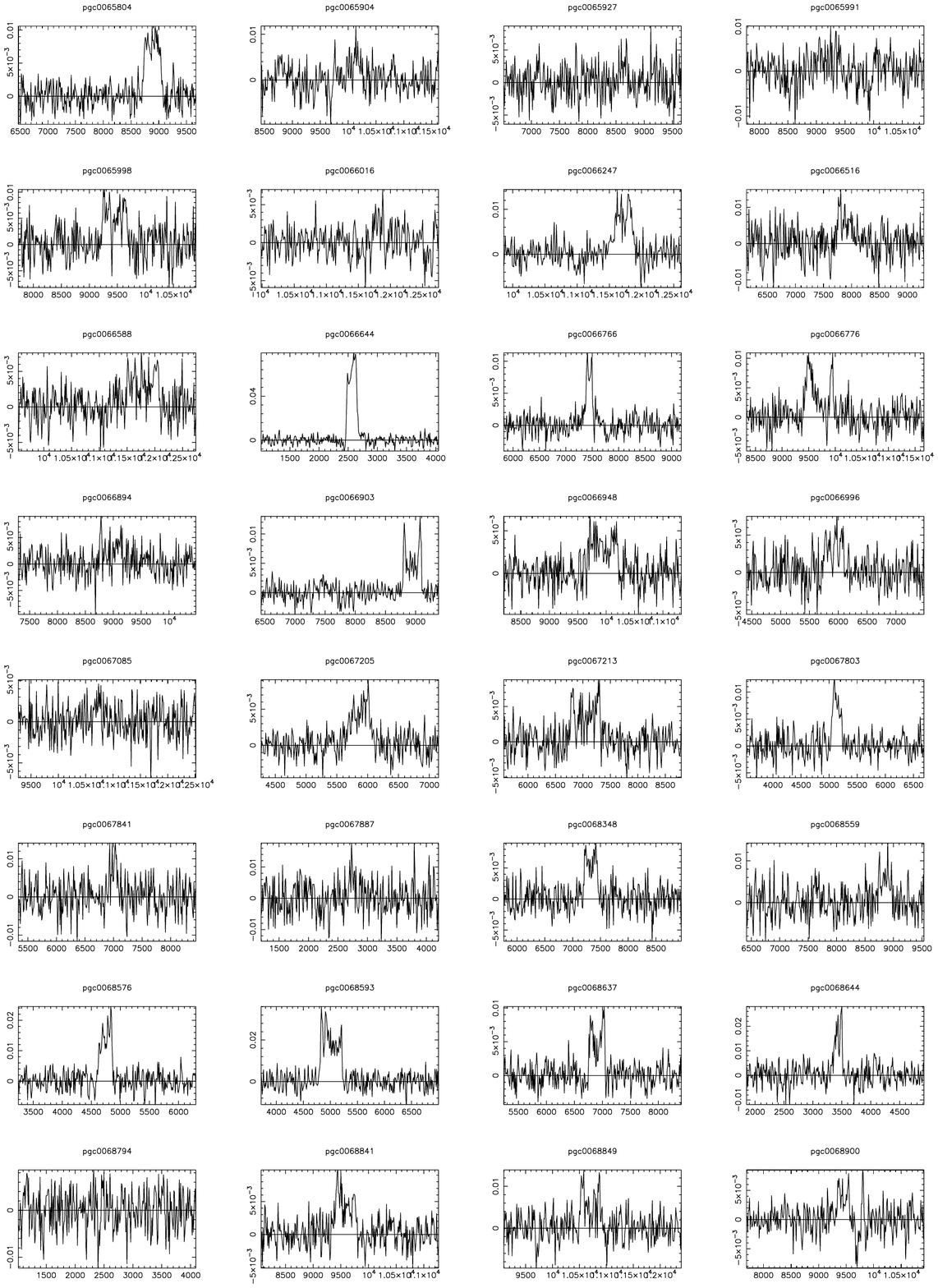}
\captcont{\small Fig. 2 {\bf o.} HI profiles. Continued.}
\end{figure*}

\clearpage
\newpage

\begin{figure*}
\centering
\includegraphics[width=\textwidth]{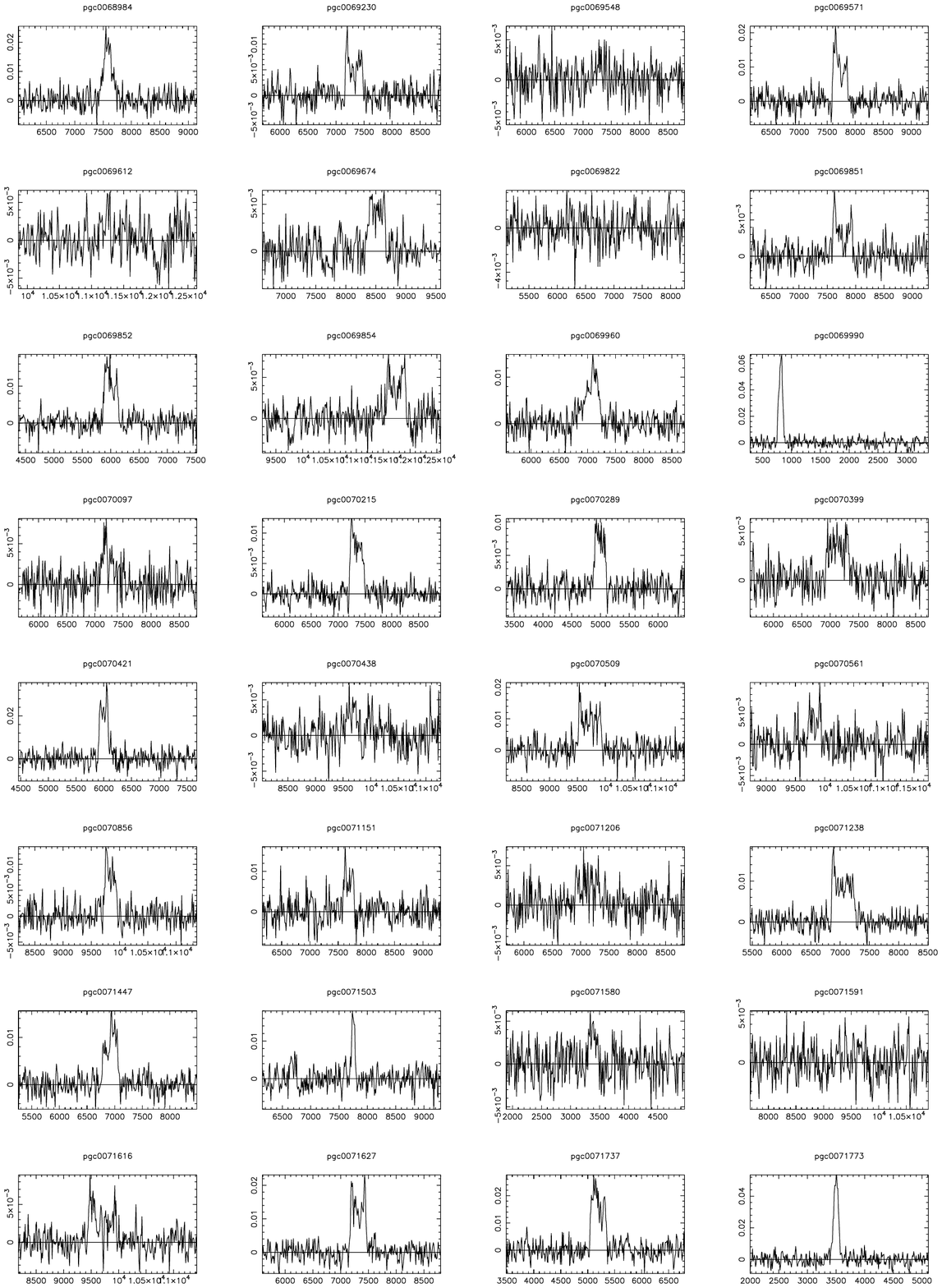}
\captcont{\small Fig. 2  {\bf p.} HI profiles. Continued.}
\end{figure*}

\clearpage
\newpage

\begin{figure*}
\centering
\includegraphics[width=\textwidth]{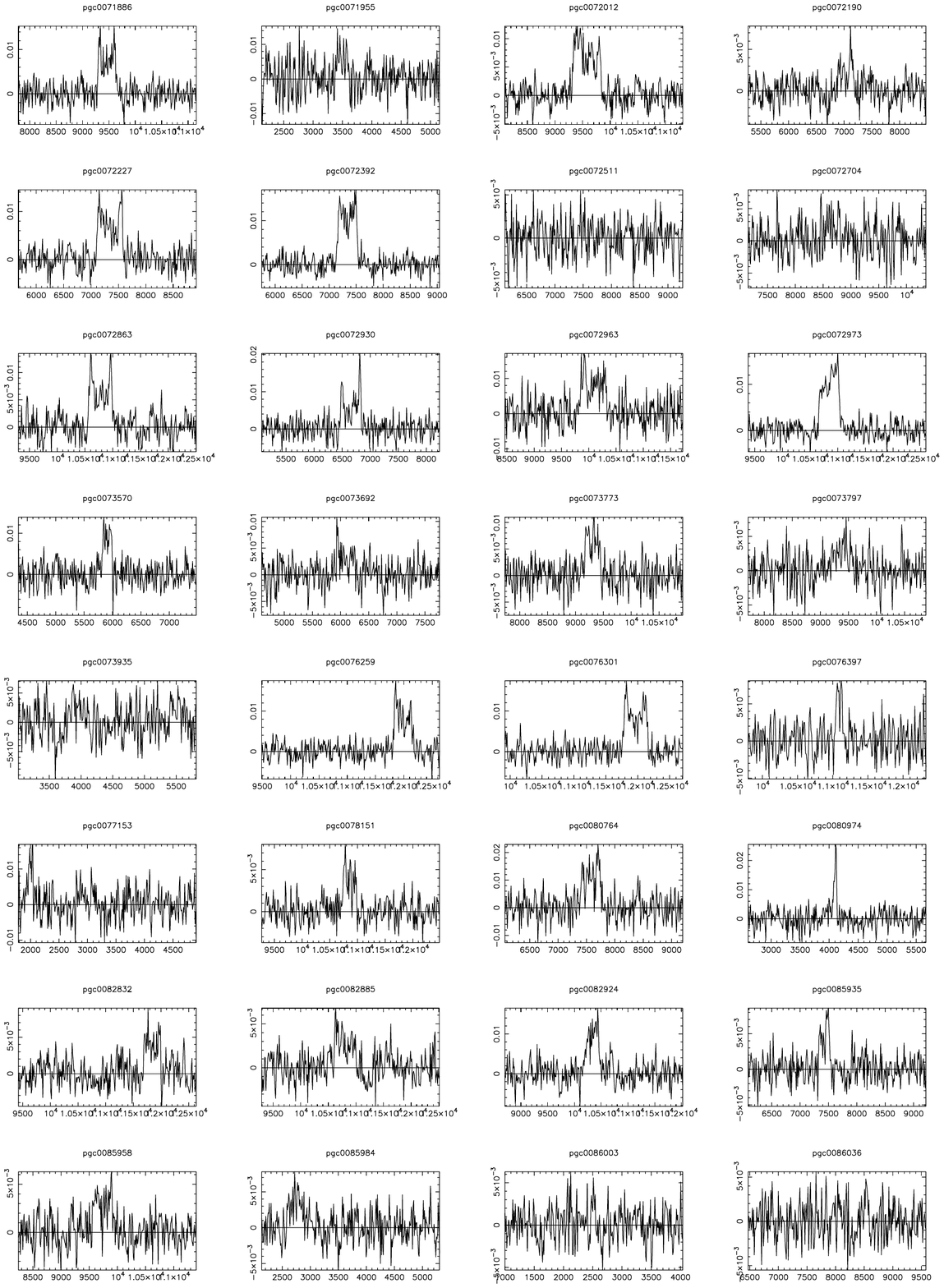}
\captcont{\small Fig. 2 {\bf q.} HI profiles. Continued.}
\end{figure*}

\clearpage
\newpage

\begin{figure*}
\centering
\includegraphics[width=\textwidth]{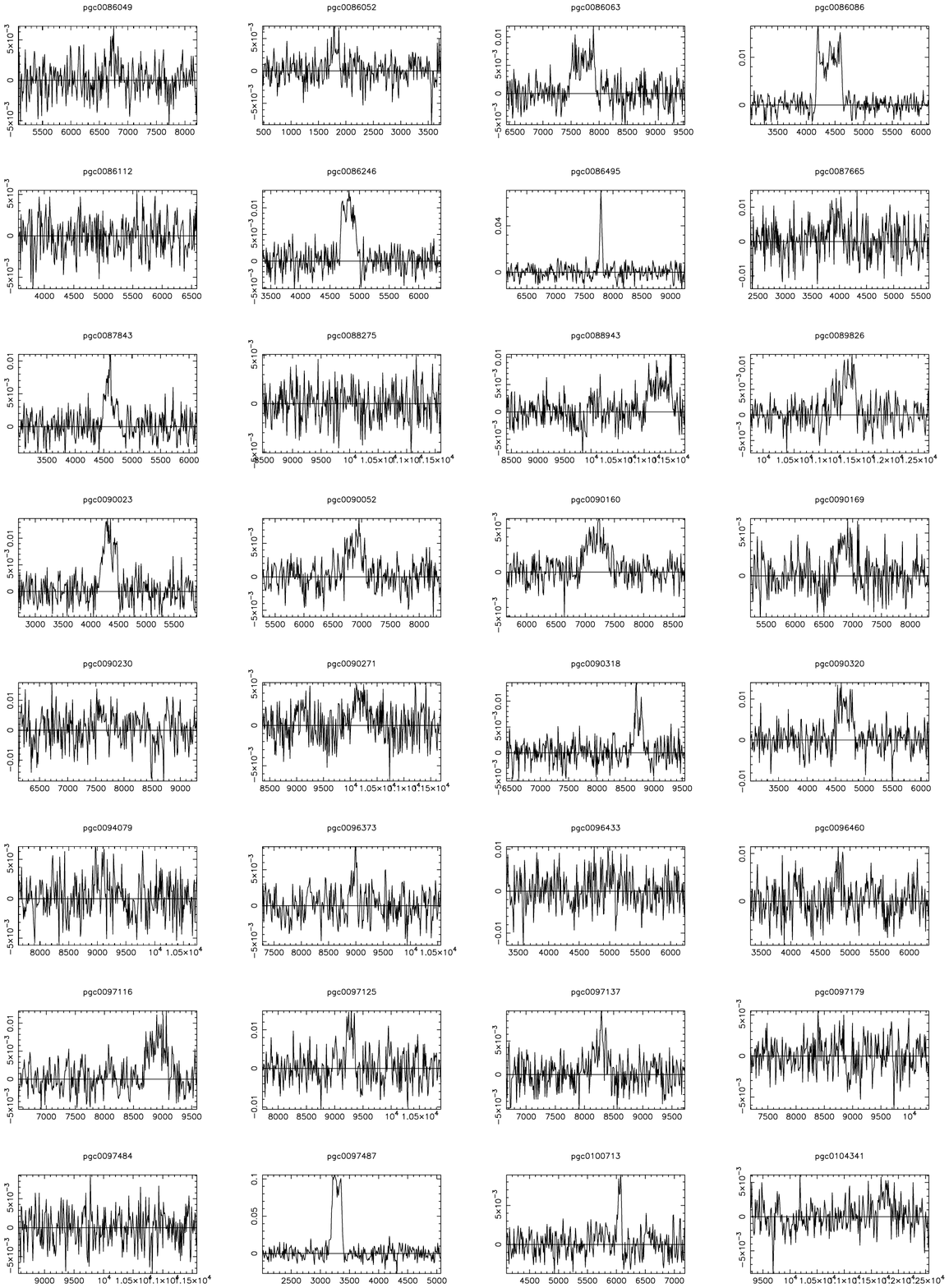}
\captcont{\small Fig. 2 {\bf r.} HI profiles. Continued.}
\end{figure*}

\clearpage
\newpage

\begin{figure*}
\centering
\includegraphics[width=\textwidth]{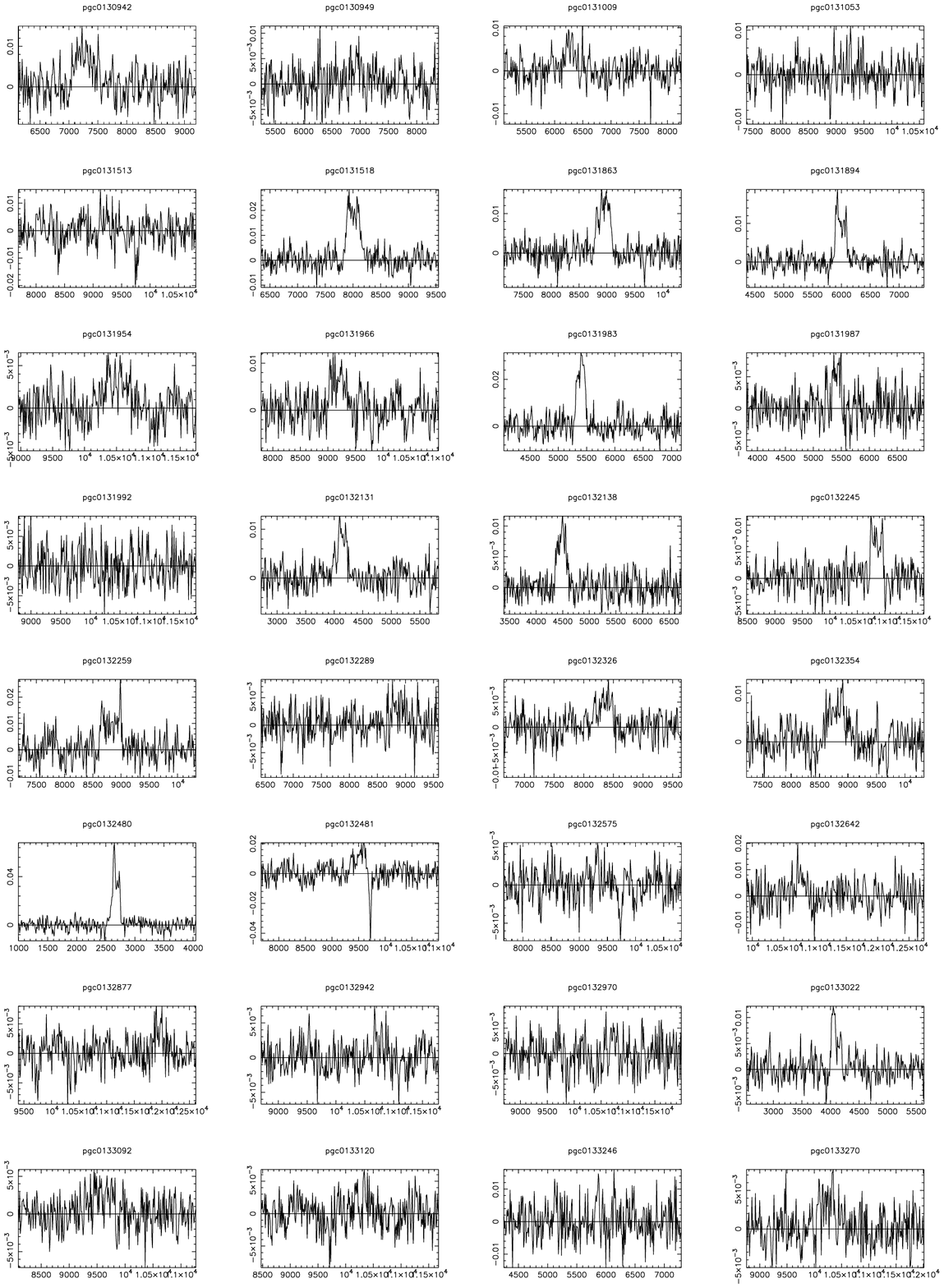}
\captcont{\small Fig. 2 {\bf s.} HI profiles. Continued.}
\end{figure*}

\clearpage
\newpage

\begin{figure*}
\centering
\includegraphics[width=\textwidth]{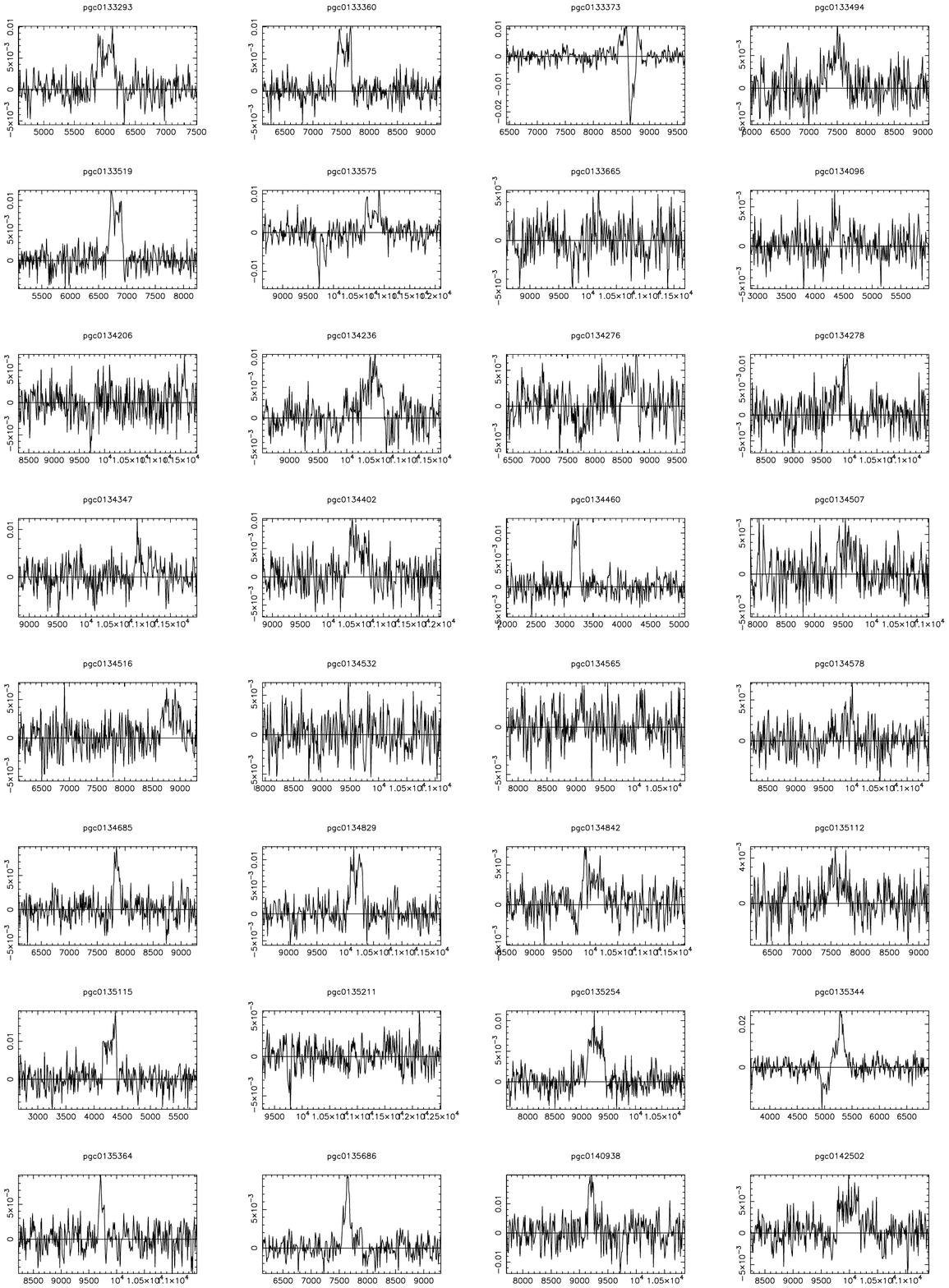}
\captcont{\small Fig. 2 {\bf t.} HI profiles. Continued.}
\end{figure*}

\clearpage
\newpage

\begin{figure*}
\centering
\includegraphics[width=\textwidth]{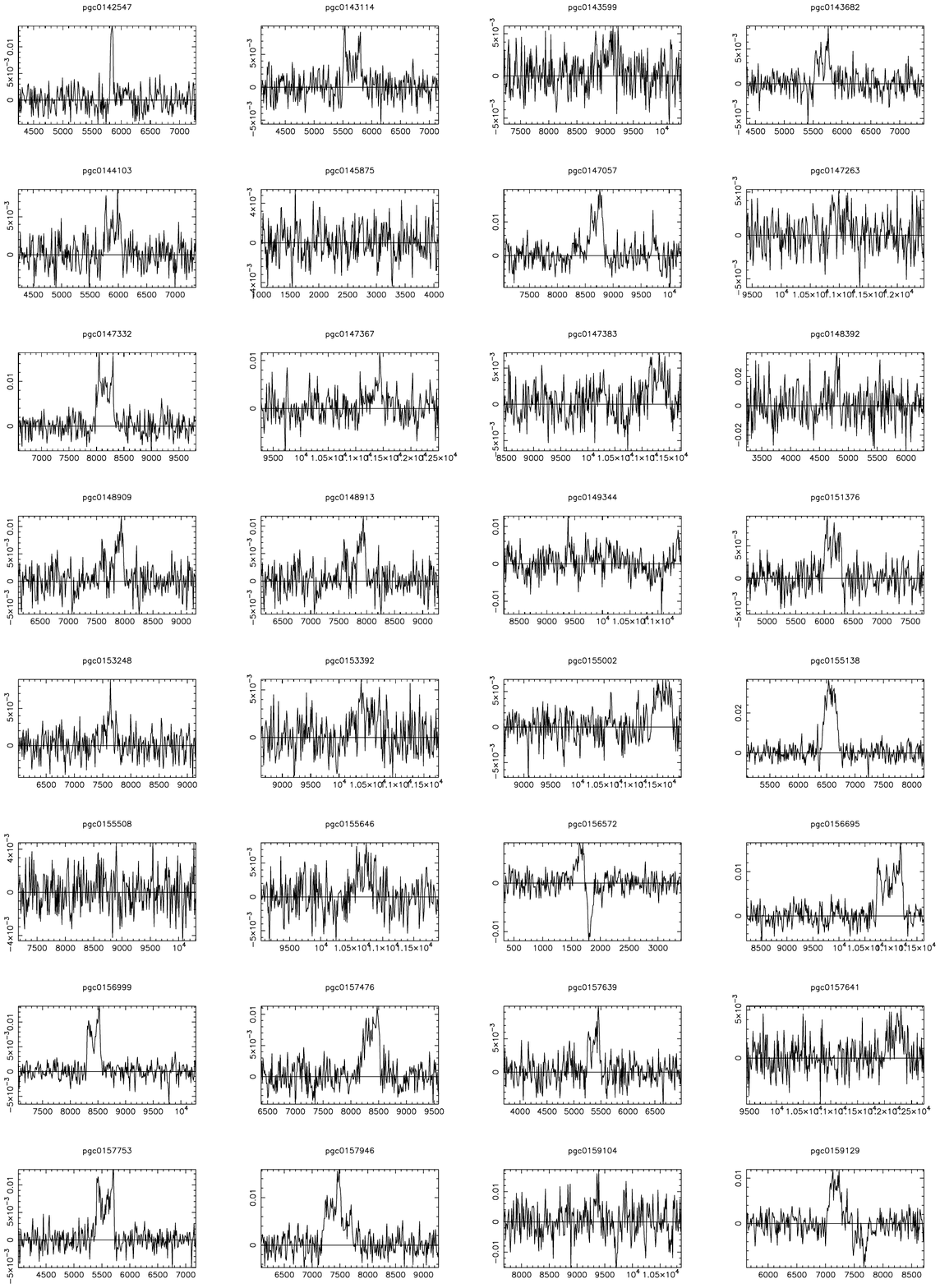}
\captcont{\small Fig. 2 {\bf u.} HI profiles. Continued.}
\end{figure*}

\clearpage
\newpage

\begin{figure*}
\centering
\includegraphics[width=\textwidth]{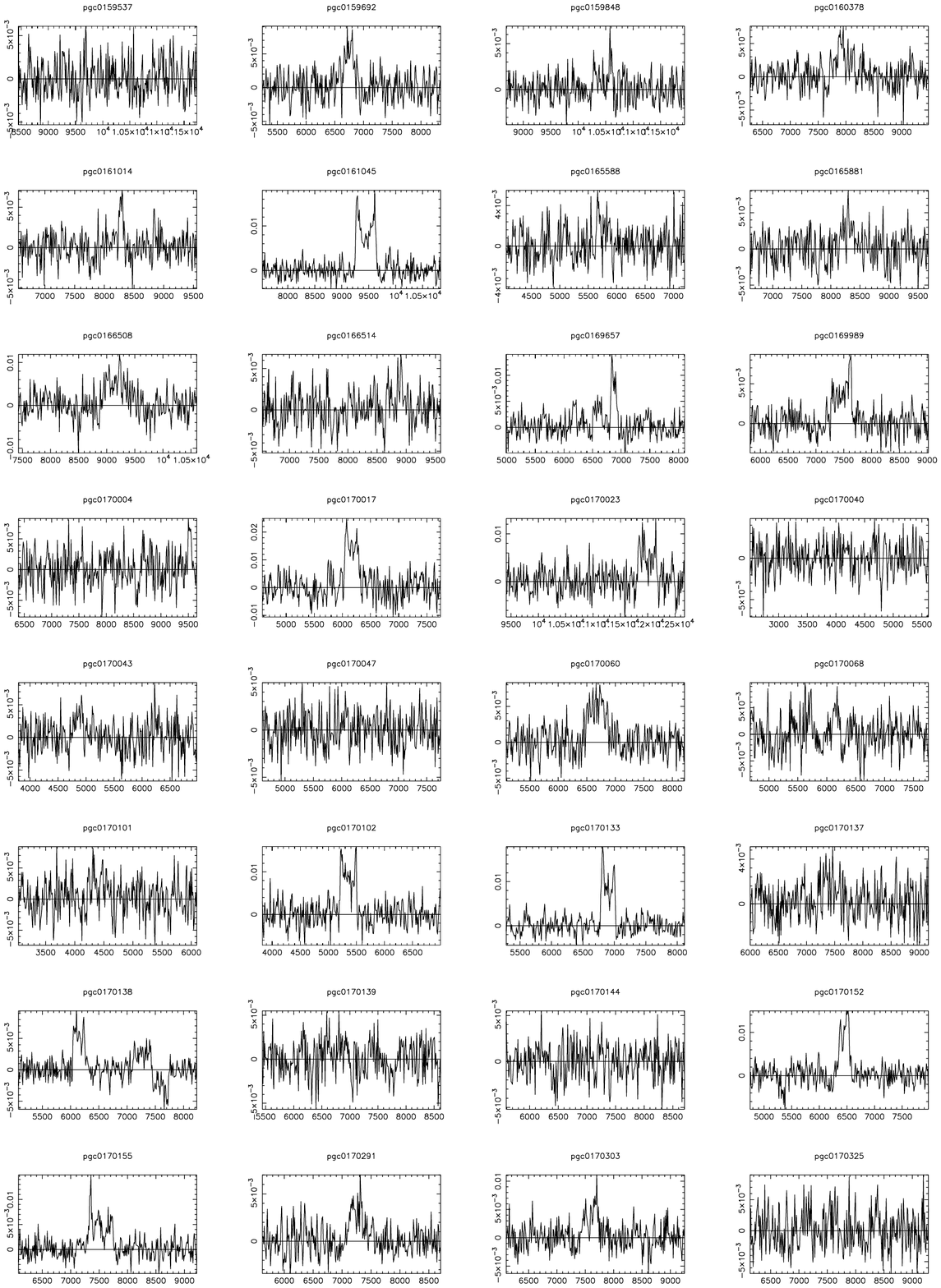}
\captcont{\small Fig. 2 {\bf v.} HI profiles. Continued.}
\end{figure*}

\clearpage
\newpage

\begin{figure*}
\centering
\includegraphics[width=\textwidth]{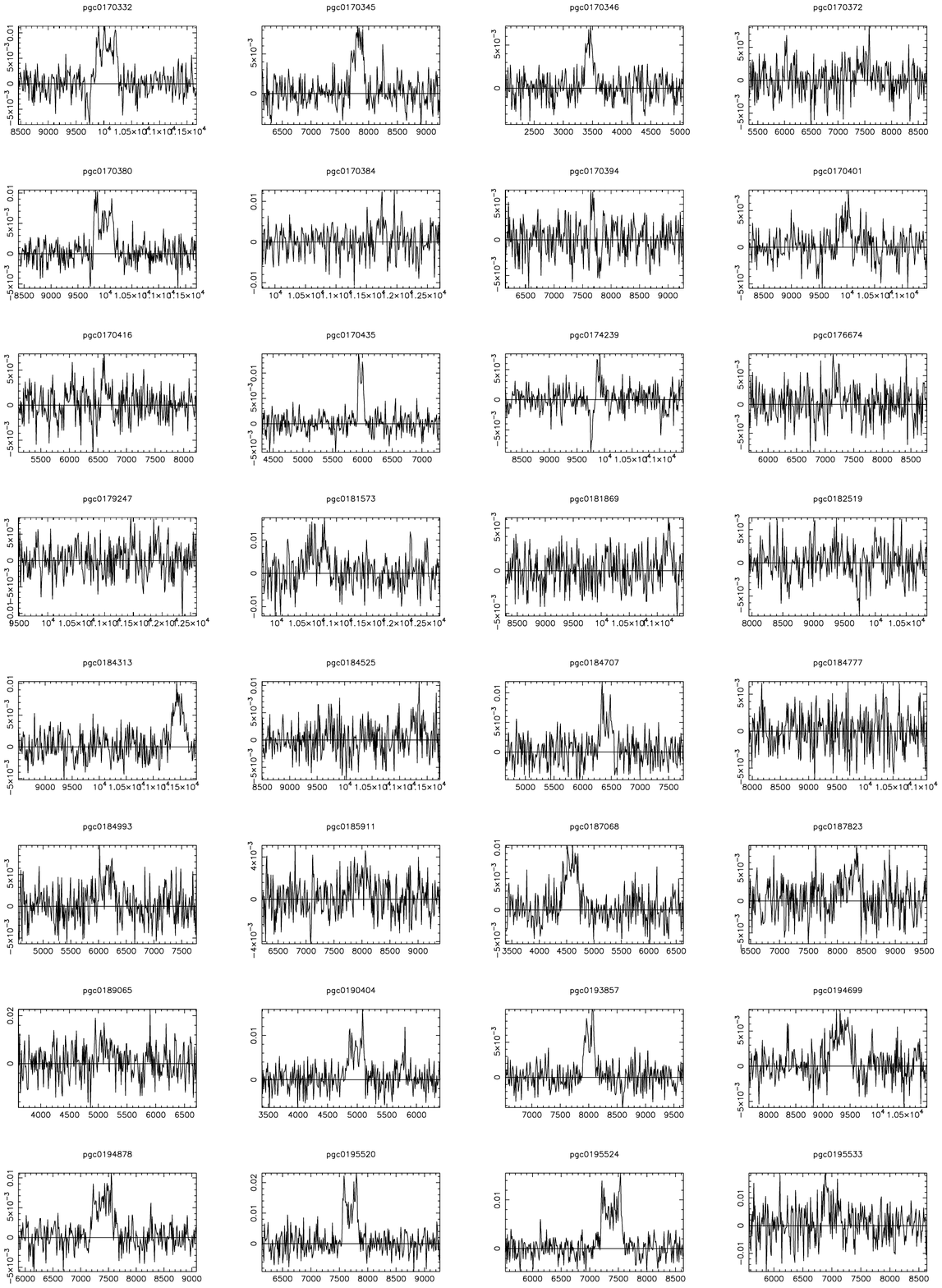}
\captcont{\small Fig. 2 {\bf w.} HI profiles. Continued.}
\end{figure*}

\clearpage
\newpage

\begin{figure*}
\centering
\includegraphics[width=\textwidth]{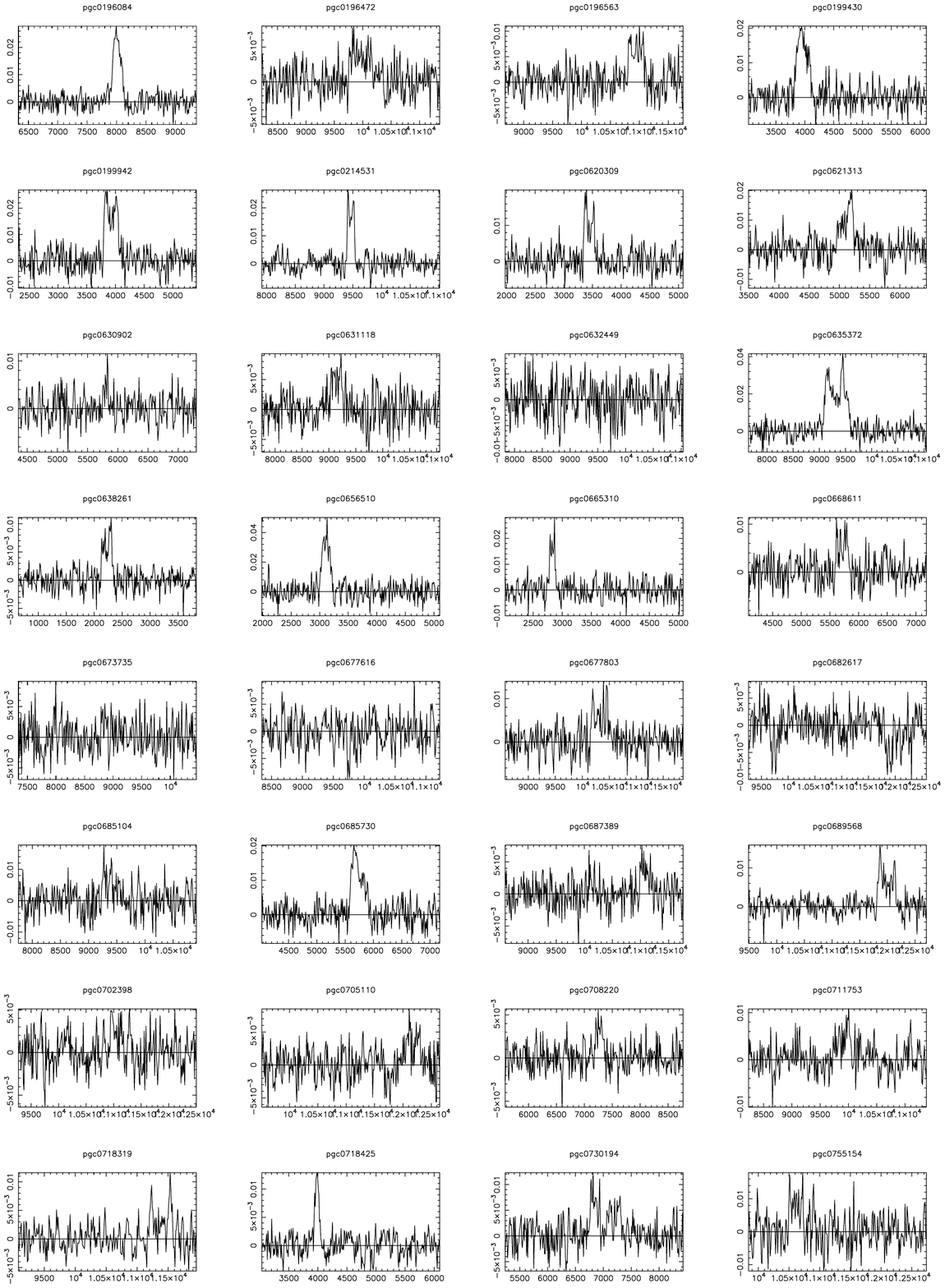}
\captcont{\small Fig. 2 {\bf x.} HI profiles. Continued.}
\end{figure*}

\clearpage
\newpage

\begin{figure*}
\centering
\includegraphics[width=\textwidth]{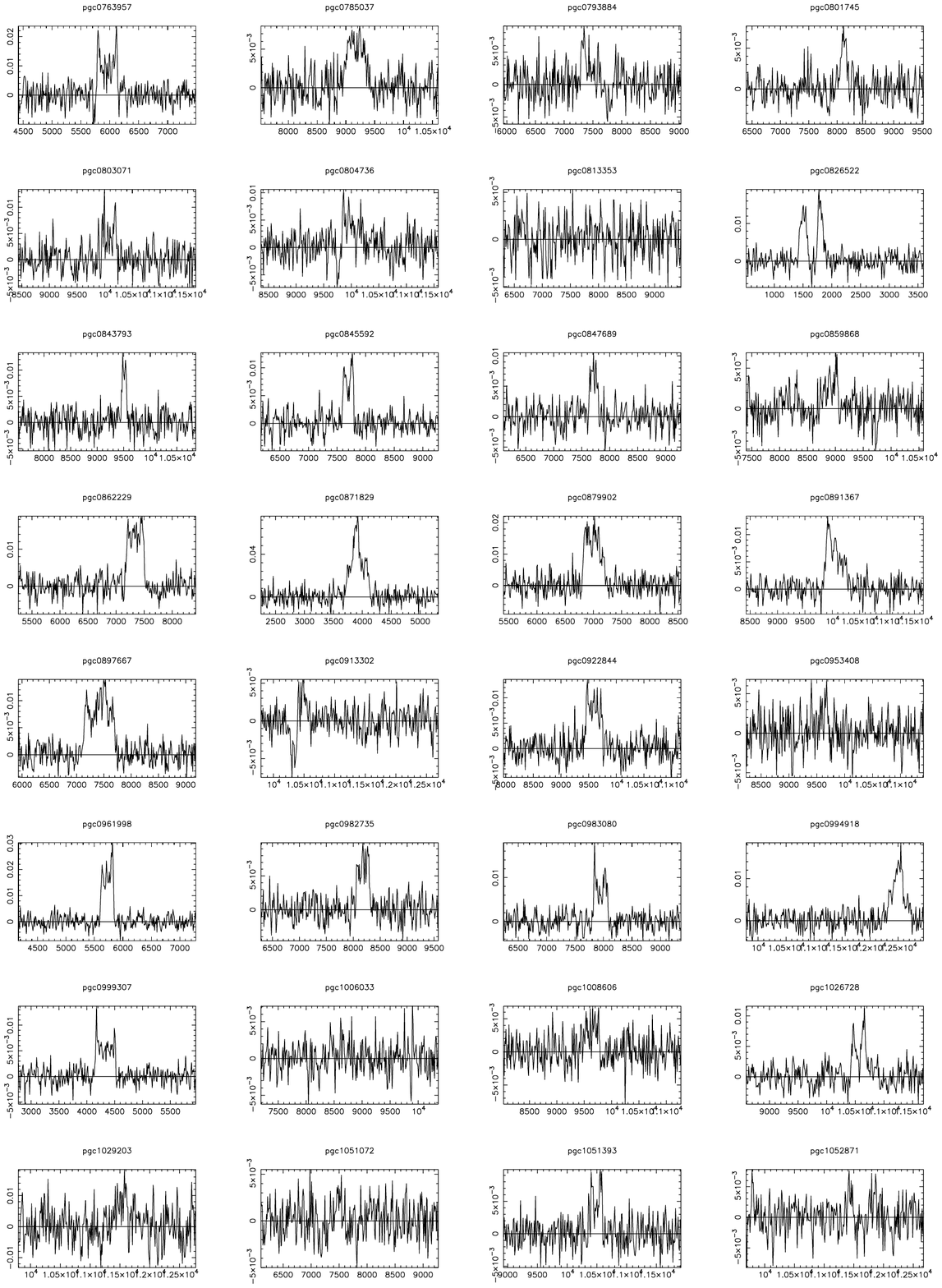}
\captcont{\small Fig. 2 {\bf y.} HI profiles. Continued.}
\end{figure*}

\clearpage
\newpage

\begin{figure*}
\centering
\includegraphics[width=\textwidth]{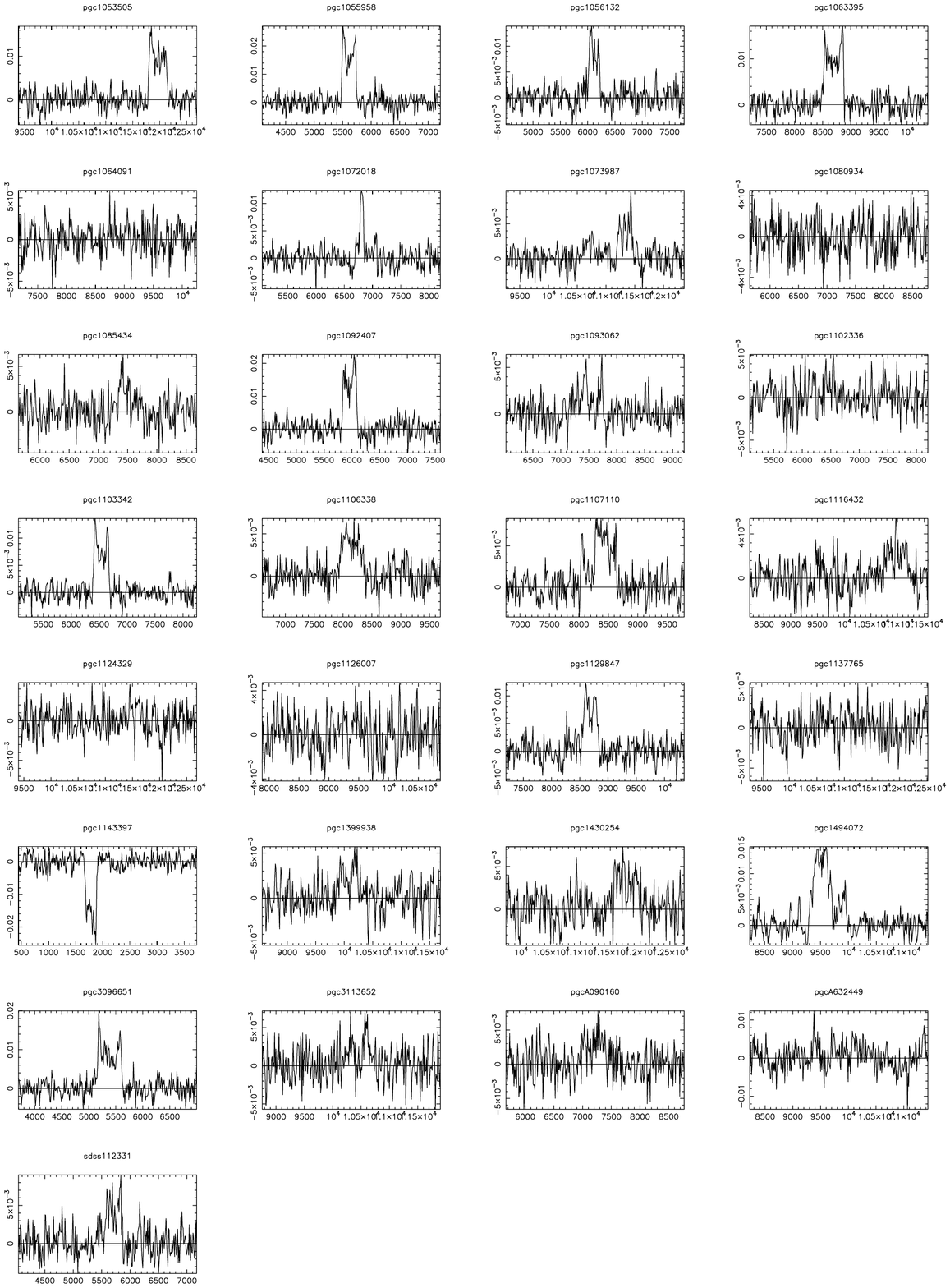}
\captcont{\small Fig. 2 {\bf z.} HI profiles. Continued.}
\end{figure*}

\end{appendix}

\end{document}